\documentclass[aps,prl,twocolumn,showpacs,superscriptaddress,floatfix,10pt,longbibliography]{revtex4-1}

\usepackage{framed}
\usepackage{graphicx}
\usepackage{amsmath}
\usepackage{lmodern}
\usepackage{epsfig}
\usepackage{helvet}
\usepackage{framed}
\usepackage{bbold}
\usepackage{subfigure}
\usepackage{hyperref}

\usepackage{bm}

\hypersetup{
  colorlinks   = true, 
  urlcolor     = blue, 
  linkcolor    = blue, 
  citecolor   = red, 
}

\newcommand{\bra}[1]{\mbox{$\langle #1 |$}}
\newcommand{\ket}[1]{\mbox{$| #1 \rangle$}}
\newcommand{\braket}[2]{\mbox{$\langle #1 | #2 \rangle$}}

\newcommand{\proj}[1]{\mbox{$| #1 \rangle\langle #1 |$} }

\newcommand{\tr}{\mbox{$\text{tr}$}}

\def\CFT{\mbox{\tiny CFT}}

\newcommand{\phdag}{{\phantom{\dagger}}}

\allowdisplaybreaks[3]

\def\appendix{Appendix }

\DeclareMathOperator{\real}{Re}

\newcommand{\beq}{\begin{equation}}
\newcommand{\eeq}{\end{equation}}
\newcommand{\beqa}{\begin{eqnarray}} 
\newcommand{\eeqa}{\end{eqnarray}}
\newcommand{\blgn}{\begin{aligned}}
\newcommand{\elgn}{\end{aligned}}

\newcommand{\la}{\langle}
\newcommand{\ra}{\rangle}
\newcommand{\da}{\dagger}

\newcommand{\f}{\frac}

\newcommand{\al}{\alpha}
\newcommand{\be}{\beta}
\newcommand{\ga}{\gamma}

\newcommand{\ep}{\epsilon}
\newcommand{\pa}{\partial}

\newcommand{\zb}{\bar z} 
\newcommand{\Tb}{\bar T} 

\allowdisplaybreaks[3]

\begin{document}

\title{Emergent universality in critical quantum spin chains:\\ entanglement Virasoro algebra}

\author{Q. Hu}
\email{qhu@perimeterinstitute.ca}
\affiliation{Perimeter Institute for Theoretical Physics, Waterloo, Ontario N2L 2Y5, Canada}  \date{\today}
\affiliation{Department of Physics and Astronomy, University of Waterloo, Waterloo, Ontario N2L 3G1,
Canada}
\author{A. Franco-Rubio}
\affiliation{Perimeter Institute for Theoretical Physics, Waterloo, Ontario N2L 2Y5, Canada}
\affiliation{Department of Physics and Astronomy, University of Waterloo, Waterloo, Ontario N2L 3G1,
Canada}
\affiliation{Max-Planck-Institut f\"ur Quantenoptik, Hans-Kopfermann-Stra\ss e 1, 85748 Garching, Germany}  
\affiliation{Munich Center for Quantum Science and Technology, Schellingstra\ss e 4, 80799 M\"unchen, Germany}

\author{G. Vidal}
\affiliation{Perimeter Institute for Theoretical Physics, Waterloo, Ontario N2L 2Y5, Canada}
\affiliation{X, The Moonshot Factory, Mountain View, CA 94043, USA}
\date{\today}

\date{\today}

\begin{abstract}

Entanglement entropy and entanglement spectrum have been widely used to characterize quantum entanglement in extended many-body systems. Given a pure state of the system and a division into regions $A$ and $B$, they can be obtained in terms of the \textit{Schmidt values}, or eigenvalues $\lambda_{\alpha}$ of the reduced density matrix $\rho_A$ for region $A$. In this paper we draw attention instead to the \textit{Schmidt vectors}, or eigenvectors $\ket{v_{\alpha}}$ of $\rho_A$. We consider the ground state of critical quantum spin chains whose low energy/long distance physics is described by an emergent conformal field theory (CFT). We show that the Schmidt vectors $\ket{v_{\alpha}}$ display an emergent universal structure, corresponding to a realization of the Virasoro algebra of a boundary CFT (a chiral version of the original CFT). Indeed, we build weighted sums $H_n$ of the lattice Hamiltonian density $h_{j,j+1}$ over region $A$ and show that the matrix elements $\bra{v_{\alpha}}H_n \ket{v_{\alpha'}}$ are universal, up to finite-size corrections.
More concretely, these matrix elements are given by an analogous expression for $H_n^{\CFT} = \f 1 2 (L_n + L_{-n})$ in the boundary CFT, where $L_n$'s are (one copy of) the Virasoro generators. 
We numerically confirm our results using the critical Ising quantum spin chain and other (free-fermion equivalent) models.
\end{abstract}

\date{\today}

\maketitle

The study of quantum entanglement has led to valuable insights into the physics of quantum many-body phenomena. For instance, both at (or near) a quantum critical point \cite{vidal2003entanglement, holzhey1994geometric, calabrese2004entanglement,lauchli2013operator,ohmori2015physics, cardy2016entanglement,roy2020entanglement,cho2017universal,calabrese2008entanglement,de2012entanglement,lepori2013scaling,giampaolo2013universal,laflorencie2014spin,chandran2014universal,lundgren2016universal,schuler2016universal,whitsitt2017spectrum,surace2019operator,eisler2020entanglement,tang2020critical} and in many-body systems with topological order \cite{li2008entanglement,thomale2010entanglement,fidkowski2010entanglement,prodan2010entanglement,turner2010entanglement,qi2012general}, entanglement has been seen to display \textit{universal} behaviour. Indeed, in a quantum critical system (or in a topologically ordered material), several entanglement properties of the ground state and low energy eigenstates are dictated by the universality class of the corresponding critical point (correspondingly, topological phase) and are thus largely independent of microscopic details. This observation has turned quantum entanglement into a useful theoretical tool to diagnose the universality class of a many-body wavefunction.

Consider a quantum system made of two parts $A$ and $B$, such that the Hilbert space can be written as the tensor product $\mathbb{V}^{AB} \cong \mathbb{V}^{A}\otimes \mathbb{V}^{B}$. The Schmidt decomposition of a pure state $\ket{\Psi} \in \mathbb{V}^{AB}$ reads
\begin{equation} \label{eq:Schmidt}
    \ket{\Psi} = \sum_{\alpha=1}^{\chi} \lambda_{\alpha} \ket{v_{\alpha}}\ket{w_{\alpha}},
\end{equation}
where $\{\lambda_{\alpha}\}$ are its Schmidt coefficients and $\{\ket{v_{\alpha}} \in \mathbb{V}^{A}\}$ and $\{\ket{w_\alpha} \in \mathbb{V}^{B}\}$ are the Schmidt vectors, with $\lambda_{\alpha} \ge \lambda_{\alpha+1} > 0$ and $\braket{v_{\alpha}}{v_{\alpha'}} = \braket{w_{\alpha}}{w_{\alpha'}} = \delta_{\alpha,\alpha'}$. Then subsystem $A$ is described by the reduced density matrix
\begin{equation} \label{eq:rhoA}
    \rho_A \equiv\tr_{B} |\Psi\rangle\langle \Psi| = \sum_{\alpha} \lambda_{\alpha}^{2} \proj{v_\alpha}.~~~~~~
\end{equation}
Historically, the \textit{entanglement entropy} \cite{bennett1996concentrating}, defined as
\begin{equation}
    S_A \equiv -\tr \left(\rho_A \log \rho_A\right) = - \sum_{\alpha} \lambda^{2}_{\alpha} \log \lambda_\al^{2},
\end{equation} 
was the first entanglement magnitude for which universal behaviour was reported.
For a critical lattice system, the entanglement entropy of the ground state wavefunction scales universally as \cite{vidal2003entanglement, holzhey1994geometric, calabrese2004entanglement}:
\begin{equation} \label{eq:SA}
    S_A\sim \frac{c}{3}\log (L/a),
\end{equation}
where $L$ is the size of part $A$, $a$ is the lattice spacing, and $c$ is the central charge of the underlying conformal field theory (CFT). Eq. \eqref{eq:SA} is also valid in the continuum for the ground state of a 1+1 dimensional CFT, with $a$ representing some UV regulator. Later on, universality was also observed in the \textit{entanglement spectrum} \cite{li2008entanglement}, defined as the eigenvalues $\{E_{\alpha}\}$ of the so-called entanglement Hamiltonian 
\begin{equation} \label{eq:KA}
    K_A \equiv -\f{1}{2\pi}\log \left( \rho_A \right) = \sum_{\alpha} E_{\alpha} \proj{v_\alpha},
\end{equation}
with $E_{\alpha} \leq E_{\alpha+1} < \infty$. For critical quantum spin chains, the entanglement spectrum $\{E_{\alpha}\}$ reproduces the spectrum of scaling dimensions $\{h_{\alpha}\}$ of a related boundary conformal field theory (BCFT) \cite{lauchli2013operator, cardy2016entanglement}. 

Recall that, in a lattice model, region $A$ is made of lattice sites (each hosting a quantum spin) and therefore the vector space $\mathbb{V}^{A}$ itself factorizes as the tensor product of vector spaces $\mathbb{V}_j$ for individual sites $j \in A$, 
\begin{equation} \label{eq:factorization}
    \mathbb{V}^{A} \cong \bigotimes_{j\in A} \mathbb{V}_j.
\end{equation}
The prevalent paradigm of bipartite entanglement is concerned only with the Schmidt coefficients $\{\lambda_{\alpha}\}$ in \eqref{eq:Schmidt} or, equivalently, the entanglement energies $\{E_{\alpha}\}$, with $E_\alpha = -\f{1}{2\pi}\log(\lambda_\alpha^2)$. One may however further inquire how this entanglement relates to factorization \eqref{eq:factorization}, thereby probing its multipartite character. For instance, the entanglement contour \cite{chen2014entanglement} and related work \cite{botero2004spatial,frerot2015area,coser2017contour,tonni2018entanglement,wen2018fine,kudler2019holographic,wen2020entanglement,han2019entanglement,ageev2019entanglement,kudler2020negativity} describe how the entanglement entropy $S_A$ of region $A$ is distributed among the sites $j \in A$.

In this paper we investigate how the Schmidt vectors $\{\ket{v_{\alpha}}\}$ are embedded in the tensor product structure \eqref{eq:factorization}. While this is in general a very difficult problem, we will see that much can be said for critical quantum spin chains. Let $H=\sum_{j} h_{j,j+1}$ be the Hamiltonian of an infinite chain, with $h_{j,j+1}$ its nearest neighbor \textit{Hamiltonian density}. We will conjecture, and numerically confirm, an emergent relation between (i) the Schmidt vectors $\{\ket{v_{\alpha}}\}$ on region $A$ of the ground state of $H$ and (ii) the Hamiltonian density $h_{j,j+1}$. In practice, it is natural to introduce certain operators $H_n$ (weighted sums of $h_{j,j+1}$ within region $A$, see Eqs. \eqref{eq:H0lat}-\eqref{eq:Hnlat}) and express this relation in terms of analogous operators $H_n^{\CFT}$ in the CFT (Eq. \eqref{eq:conjecture2}). It then becomes manifest that this relation is \textit{universal}, in that the operators $H_n^{\CFT}$ only depend on the universality class of the critical point, as expressed by the corresponding BCFT. From this relation we then learn that the Schmidt vectors are embedded in the tensor product structure \eqref{eq:factorization} in a highly fine-tuned way, one that e.g. allows us to manipulate them in a controlled, well-understood manner by means of the operators $H_n$. 

We start by discussing CFT facts in the continuum that motivate our proposal on the lattice.

\textit{Entanglement Virasoro algebra in the continuum.---} Consider the ground state $\ket{\Psi^{\CFT}}$ of a 1+1 dimensional CFT Hamiltonian $H^{\CFT} = \int_{-\infty}^{\infty} dx~ h(x)$ and a finite interval $(-R,R)$. To have a decomposition of the total Hilbert space, one needs to specify the boundary condition $\ga$ imposed on two small spatial regions of thickness $\ep$ around the entanglement cuts $x=\pm R$ \cite{ohmori2015physics,cardy2016entanglement}:
\begin{equation}
    \mathbb{V}\to \mathbb{V}^{A,\ga}\otimes \mathbb{V}^{B,\ga},
\end{equation}
where $A=(-R+\ep,R-\ep)$ and $B=(-\infty,-R-\ep)\cup(R+\ep,\infty)$. It is well-known \cite{casini2011towards,hislop1982modular} that the entanglement Hamiltonian $K^{\CFT}_{A}$ can be expressed as an integral of the Hamiltonian density $h(x)$,
\begin{equation} \label{eq:KACFT}
K^{\CFT}_{A}=\int_{-R+\epsilon}^{R-\epsilon}dx\frac{R^2-x^2}{2R}h(x) + a_1,
\end{equation}
where the constant $a_1$ enforces $\tr \left(\rho_{A}^{\CFT}\right) = 1$. Here we point out (see Appendix I for a derivation) that one can similarly obtain an entire representation of the Virasoro algebra $\{L_n\}_{n\in \mathbb{Z}}$ \cite{francesco2012conformal},
\begin{equation}
    [L_n, L_m] = (n-m)L_{n+m} + \frac{c}{12}n(n^2-1) \delta_{m+n,0},
    \label{eq:VirasoroAlgebra}
\end{equation}
acting on the Schmidt vectors $\{\ket{v^{\CFT}_{\alpha}}\}$ of region $A$. For convenience, we work with $H_{n} \equiv \frac 1 2(L_{n} + L_{-n})$ for $n \geq 0$, which can be expressed in terms of $h(x)$ as
\begin{equation}
H_0^{\CFT} = \dfrac{l}{\pi}\int_{-R+\epsilon}^{R-\epsilon}\!\!dx\,\frac{R^2-x^2}{R}h(x)+ \frac{c}{24}\left(1 + \frac{4l}{\pi^2}\right),~~~\label{eq:H0CFT}
\end{equation}
and
\begin{equation}
H_n^{\CFT} =\dfrac{l}{\pi}\int_{-R+\epsilon}^{R-\epsilon} dx\,\frac{R^2-x^2}{R} \cos(n\theta(x))h(x)~~~~~~~\label{eq:HnCFT}
\end{equation}
for $n\neq 0$. Here we have defined $l \equiv \log\left(\f{2R}{\ep}\right)$ and $\theta(x) \equiv \frac{\pi}{2} - \frac{\pi}{2l} \log\left(\frac{R+x}{R-x} \right)$. For $n=0$, $H_0^{\CFT}$ is diagonal in the Schmidt basis $\{\ket{v^{\CFT}_{\alpha}}\}$,
\begin{equation}
    H_0^{\CFT} \ket{v_{\alpha}^{\CFT}} = h^{\CFT}_{\alpha} \ket{v_{\alpha}^{\CFT}},
\end{equation}
where $\{h^{\CFT}_{\alpha}\}$ are the scaling dimensions of the BCFT. Comparing Eqs. \eqref{eq:KACFT} and \eqref{eq:H0CFT} we see that the entanglement Hamiltonian $K_A^{\CFT}$ is essentially equivalent to $H_0^{\CFT}$, 
\begin{equation} \label{eq:ESCFT}
    K_A^{\CFT} = \frac{\pi}{2l}H_0^{\CFT} + a_2,~~~E^{\CFT}_{\alpha} = \frac{\pi}{2l}h_\alpha^{\CFT} + a_2,~~~
\end{equation}
where $a_2=a_2(l)$ is a constant. For $n > 0$, the operator $H_n^{\CFT}$ can be seen to act non-diagonally, connecting different Schmidt vectors within a \textit{conformal tower} of the BCFT. 

\begin{figure}[h]
	\centering
	\includegraphics[width=0.6\linewidth]{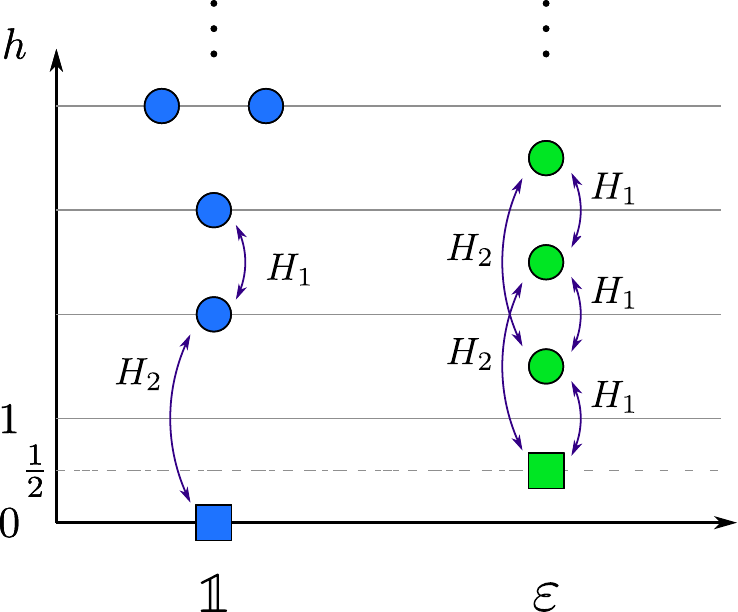}
	\caption{Conformal towers of the Ising CFT with Neumann boundary conditions.
	}
	\label{fig:Ising_towers}
\end{figure}

To illustrate the above, and for later reference, let us project the above operators onto the few first Schmidt vectors $\{\ket{v_{\alpha}^{\CFT}}\}$. In the Ising CFT (for an interval with Neumann boundary conditions), the diagonal of $H^{\CFT}_{0}$ has the scaling dimensions (see Fig. \ref{fig:Ising_towers})
\begin{equation} \label{eq:hIsingCFT}
h^{\CFT}_{\alpha} = \left\{0, 0.5, 1.5, 2, 2.5, 3, 3.5,4,4,\cdots \right\},
\end{equation}
whereas $H^{\CFT}_{1}$ and $H^{\CFT}_{2}$ for the seven first Schmidt vectors (Appendix II) read
\begin{equation} \label{eq:IsingCFTmain}
    \left(\begin{array}{ccccccc}
        0 & 0 & 0 & 0 & 0 & 0 & 0 \\
        0 & 0 & \f 1 2 & 0 & 0 & 0 & 0 \\
        0 & \f 1 2 & 0 & 0 & 1 & 0 & 0 \\          
        0 & 0 & 0 & 0 & 0 & 1 & 0\\
        0 & 0 & 1 & 0 & 0 & 0 & \f 3 2\\
        0 & 0 & 0 & 1 & 0 & 0 & 0\\
        0 & 0 & 0 & 0 & \f 3 2 & 0 & 0
    \end{array} \right), ~~~ \left(\begin{array}{ccccccc}
        0 & 0 & 0 & \frac{1}{4} & 0 & 0 & 0 \\
        0 & 0 & 0 & 0 & \frac{3}{4} & 0 & 0 \\
        0 & 0 & 0 & 0 & 0 & 0 & \frac{5}{4} \\          
        \frac{1}{4} & 0 & 0 & 0 & 0 & 0 & 0\\
        0 & \frac{3}{4} & 0 & 0 & 0 & 0 & 0\\
        0 & 0 & 0 & 0 & 0 & 0 & 0\\
        0 & 0 & \frac{5}{4} & 0 & 0 & 0 & 0
    \end{array} \right).
\end{equation}

\begin{figure}
	\centering
	\includegraphics[width=0.8\linewidth]{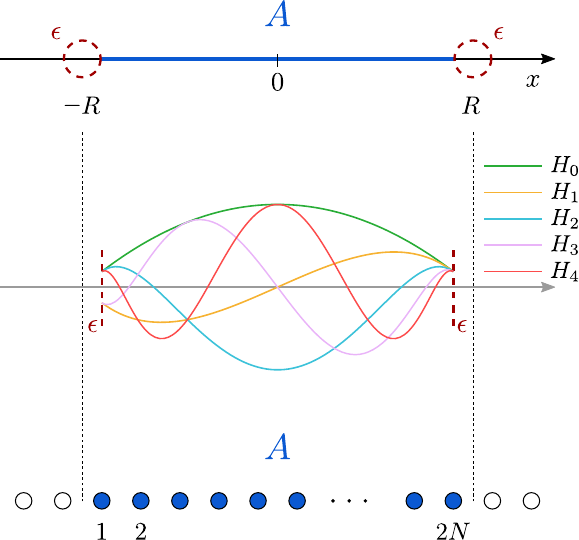}
	\caption{Region $A$ corresponds to the interval $[-R+\epsilon, R-\epsilon]$ of the real line in the CFT, and a set of $2N$ contiguous sites in the quantum spin chain. We have plotted the profiles for the first three $H_n$ operators.
	}
	\label{fig:interval}
\end{figure}

\textit{Entanglement Virasoro algebra on the lattice.---} Consider now the ground state $\ket{\Psi}$ of a critical quantum spin chain Hamiltonian $H = \sum_{j=-\infty}^{\infty} h_{j,j+1}$ on an infinite lattice, with unit lattice spacing. Let $A$ denote a finite interval made of $2N$ sites (see Fig. \ref{fig:interval}), with density matrix $\rho_A$ and entanglement Hamiltonian $K_A$ defined in Eqs. \eqref{eq:rhoA} and \eqref{eq:KA}. Assume that the critical system belongs to a universality class described by a CFT. The universality of the entanglement spectrum mentioned earlier amounts to the following conjecture from Ref. \cite{lauchli2013operator}.

\textit{Conjecture 1} (\cite{lauchli2013operator}): The entanglement spectrum $\{E_{\alpha}\}$ (eigenvalues of $K_A$) on the lattice is approximately given in terms of the entanglement spectrum $\{E_{\alpha}^{\CFT}\}$ (eigenvalues of $K_A^{\CFT}$) of the underlying BCFT with appropriate boundary conditions, up to multiplicative and additive constants. Using the RHS of Eq. \eqref{eq:ESCFT}, this reads
\begin{equation} \label{eq:Ealphah}
    E_{\alpha} \approx \frac{\pi}{2l} h_{\alpha}^{\CFT} + a_3,
\end{equation}
where $l=\log\left(\f{2N}{\ep}\right)$ depends on some UV parameter $\epsilon$ and the constant $a_3$ enforces $\tr \left( \rho_A \right)= \sum_{\alpha} e^{-2\pi E_{\alpha}}=1$. 

\begin{figure}
	\centering
    \includegraphics[width=0.5\linewidth]{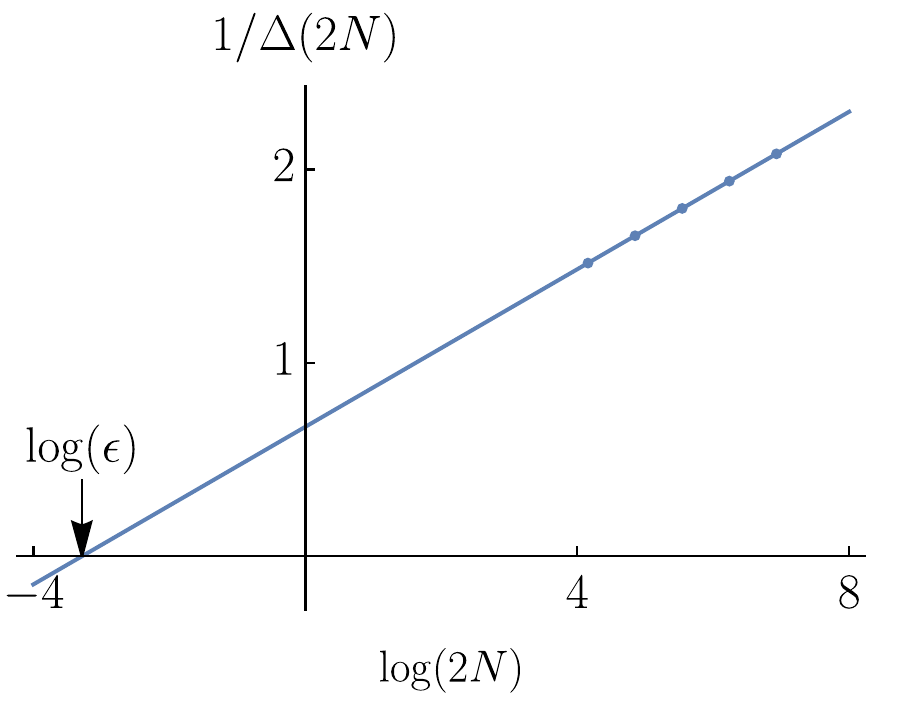}%
	\includegraphics[width=0.5\linewidth]{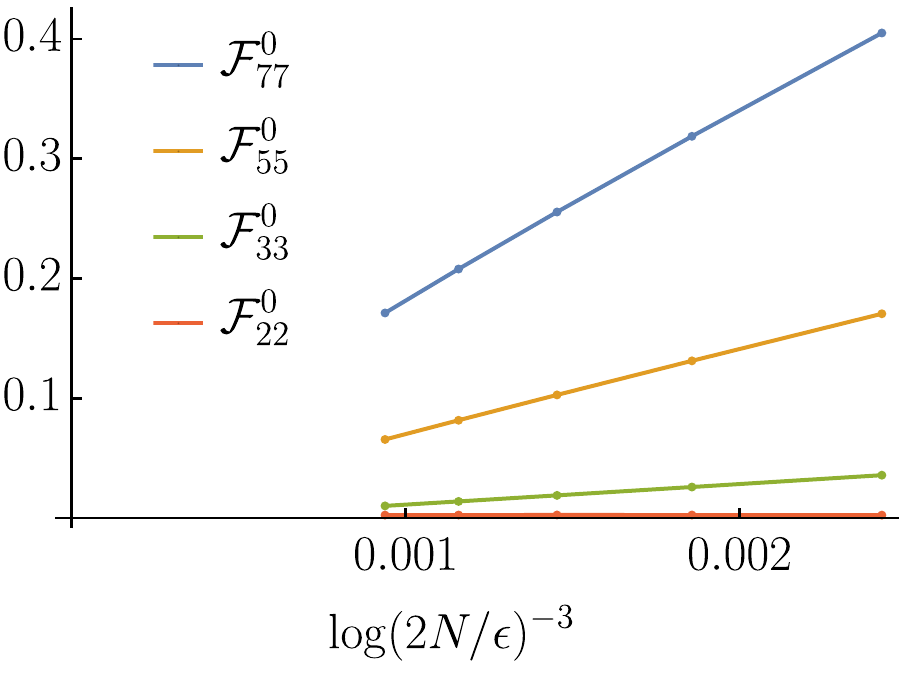}\vspace{4pt}
	\includegraphics[width=0.5\linewidth]{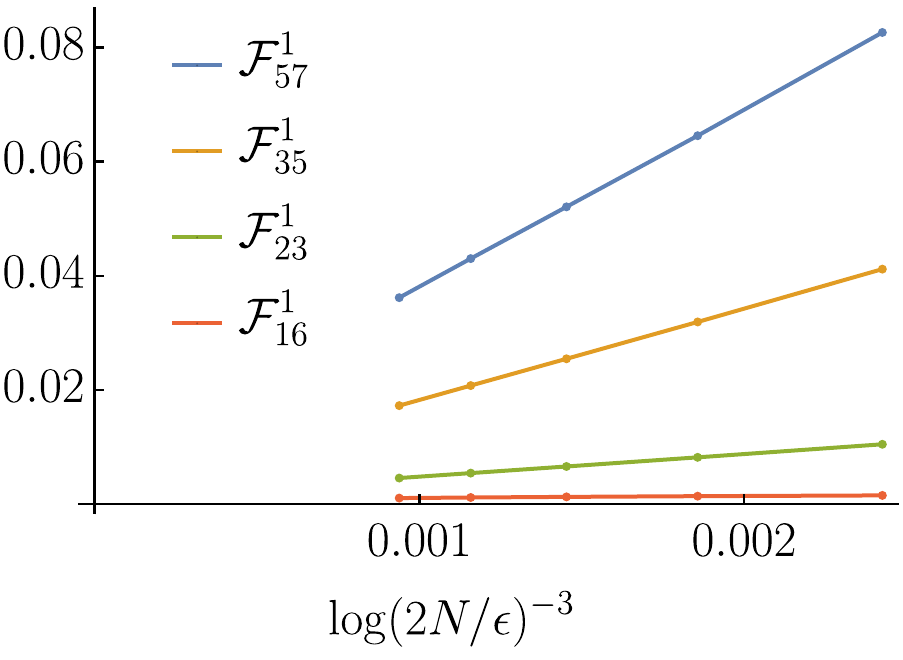}%
	\includegraphics[width=0.5\linewidth]{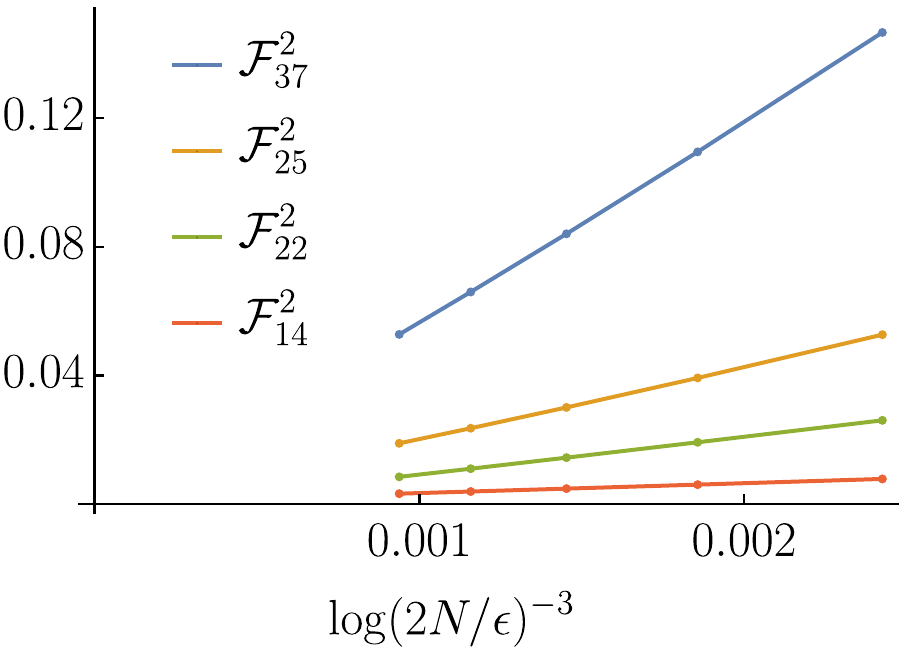}
	\caption{Numerical results for the critical quantum Ising model. 
	(Top left) Estimate of $\epsilon$ by linear extrapolation.
	(Top right) Finite size corrections $\mathcal{F}^0_{\alpha\alpha'}$ for the matrix elements of the lattice operator $H_0$. 
	(Bottom) Finite size corrections $\mathcal{F}^{1}_{\alpha\alpha'}$, $\mathcal{F}^{2}_{\alpha\alpha'}$  for the matrix elements of the lattice operators $H_{1}$, $H_2$.
	}
	\label{fig:numerics}
\end{figure}

In the above expression, $l$ is determined numerically by considering regions $A$ of different sizes. From the gap $\Delta(2N) \equiv E_2-E_1 \propto 1/l$, we can plot $1/\Delta(2N) \propto \log(2N) - \log(\epsilon)$ as a function of $\log(2N)$ for various values of $2N$ and determine $\epsilon$ by linear extrapolation, see Fig. \ref{fig:numerics}. Then we can obtain approximate values of the scaling dimensions from the lattice:
\begin{equation}
    h_\al\equiv \f{2l}{\pi}(E_\al-E_1)\approx h_\al^{\CFT}.
\end{equation}
In practice, approximation \eqref{eq:Ealphah} is seen to be more accurate for small $\alpha$ (low entanglement energy) and to be affected by finite size corrections that decay slowly as powers of $1/\log(2N/\ep)$. 

To go beyond \eqref{eq:Ealphah}, we define lattice operators 
\begin{eqnarray}
H_0 &\equiv& \frac{l}{\pi} \sum_{j=1}^{2N-1}\frac{N^{2} -x_{j+\f 1 2}^2}{N}  h_{j,j+1} + \frac{c}{24}\left(1 + \frac{4l}{\pi^2}\right),~~~\label{eq:H0lat}\\
H_n &\equiv& \frac{l}{\pi} \sum_{j=1}^{2N-1} \frac{ N^{2} -x_{j+\f 1 2}^2}{N}\cos\left(n\theta (x_{j+\f 1 2})\right)h_{j,j+1},~~~
\label{eq:Hnlat}
\end{eqnarray}
where $H_n$ is obtained from $H^{\CFT}_n$ in Eqs. \eqref{eq:H0CFT}-\eqref{eq:HnCFT} by replacing the integral with a sum over sites, $R$ with $N$, and the continuous Hamiltonian density $h(x)$ with the lattice operator $h_{j,j+1}$, which is assigned the position $x_{j+\frac{1}{2}} \equiv j -N$ (position assignment can be further refined in some cases). Our main result is the following conjecture, also relating the lattice and the continuum.
  
\textit{Conjecture 2:} Operators $\{H_n\}$ act on the Schmidt basis $\{\ket{v_\alpha}\}$ on the lattice approximately in the same way as operators $H_n^{\CFT}$ act on the Schmidt basis $\{\ket{v_\alpha^{\CFT}}\}$ in the underlying CFT (namely as a representation of the Virasoro algebra for the relevant BCFT). That is,
\begin{equation} \label{eq:conjecture2}
    \bra{v_\alpha} H_n \ket{v_{\alpha'}} \approx \bra{v^{\CFT}_\alpha} H^{\CFT}_n \ket{v^{\CFT}_{\alpha'}}.
\end{equation} 
We expect \eqref{eq:conjecture2} to again be affected by finite size corrections, and thus be more accurate for small $\alpha$ (low entanglement energy) and small $n$. 

Notice that Conjecture 1 (\cite{lauchli2013operator}) refers to the eigenvalues $\{E_{\alpha}\}$ of $K_A$, whereas Conjecture 2 involves instead its eigenvectors $\{\ket{v_{\alpha}}\}$, together with the weighted sums $H_n$ of the Hamiltonian density $h_{j,j+1}$ on region $A$. In particular, a combination of Conjecture 1 and Conjecture 2 (for $n=0$) says that the entanglement Hamiltonian $K_A$ should be approximately equivalent to $H_0$ \cite{giudici2018entanglement}
\begin{equation} \label{eq:KAH0}
    K_A  \approx \f{\pi}{2l}H_0 + a_4,
\end{equation}
where $a_4$ is a normalization constant. Notice that \eqref{eq:KAH0} is the approximate lattice version of the LHS of \eqref{eq:ESCFT}.

\textit{Example: XY model with a transverse field.---}  Consider the quantum spin Hamiltonian on an infinite chain,
\begin{equation}
    H^{XY}
    = - \frac 1 2\sum_{j=-\infty}^{\infty} \left(\frac{1+\gamma}{2}\sigma^x_{j}\sigma^x_{j+1} +  \frac{1-\gamma}{2}\sigma^y_{j}\sigma^y_{j+1} + \lambda \sigma^z_j\right),
\end{equation}
where $\sigma^{x,y,z}_{j}$'s are Pauli matrices, and $\lambda$, $\gamma$ parametrize the strength of the field and the anisotropy in the spin-spin interaction, respectively. For $\lambda=1$ and $\gamma \neq 0$, the spin chain is critical and falls within the Ising CFT universality class. For $0\leq\lambda<1$ and $\gamma = 0$, the spin chain is also critical, but within the universality class of the free Boson CFT (see Appendix III).

In particular, for $(\lambda,\gamma)=(1,1)$ we recover the \textit{critical quantum Ising model}. Using the free fermion formalism (Appendix III) we computed the lower part of the entanglement spectrum ${E_\al}$ (up to an additive constant) for regions $A$ of size $2N = \{64, 128, 256, 512, 1024\}$. Based on the entanglement spectrum we estimated $\epsilon \approx 0.037$, see Fig. \ref{fig:numerics}. The first few entries of the energy spectrum $\{h_\al\}$ for $2N=1024$ are given by
\begin{equation}
\{0, 0.50, 1.51, 2.00, 2.54, 3.04, 3.61, 4.04, 4.10, \cdots\},
\end{equation}
which match the lower part of the exact spectrum \eqref{eq:hIsingCFT} of the Ising CFT on a finite interval with Neumann boundary conditions (Cardy state $|\sigma\rangle$) as shown in Fig. \ref{fig:Ising_towers}. It is also consistent with the numerical result of \cite{ohmori2015physics}.

Finally, using the Schmidt vectors $\{\ket{v_{\alpha}}\}$, we evaluated matrix elements $\bra{v_{\alpha}} H_n \ket{v_{\alpha'}}$. For $n=0,1,2$ and $\alpha,\alpha' = 1, \cdots, 7$ we found the general form
\begin{equation} \label{eq:numerics}
    \bra{v_{\alpha}} H_n \ket{v_{\alpha'}} = \bra{v^{\CFT}_{\alpha}} H^{\CFT}_n \ket{v^{\CFT}_{\alpha'}} + \mathcal{F}^n_{\alpha\alpha'},~~~~
\end{equation}
where $\bra{v^{\CFT}_{\alpha}} H^{\CFT}_n \ket{v^{\CFT}_{\alpha'}}$ is a coefficient of one of the three matrices described in Eqs. \eqref{eq:hIsingCFT}-\eqref{eq:IsingCFTmain} and $\mathcal{F}^n_{\alpha\alpha'}$ is a finite-size correction, seen to be either zero or small. This fully agrees with \eqref{eq:conjecture2}, numerically confirming Conjecture 2.

The non-vanishing coefficients $\mathcal{F}^n_{\alpha\alpha'} \neq 0$ are seen to obey the following two general rules. (i) For fixed size of region $A$ (fixed $2N$), $|\mathcal{F}^n_{\alpha\alpha'}|$ generally grows with $\alpha,\alpha'$, so that \eqref{eq:conjecture2} is more accurate for Schmidt vectors $\ket{v_{\alpha}}$ corresponding to small entanglement energy $E_{\alpha}$. Moreover, (ii) for fixed $\alpha,\alpha'$ and $n$, coefficient $\mathcal{F}^n_{\alpha\alpha'}$ decays with $2N$ as a power of $1/\log(2N/\ep)$, as illustrated in Fig. \ref{fig:numerics}.

\textit{Universality.--} We emphasize that  $\bra{v^{\CFT}_{\alpha}} H^{\CFT}_n \ket{v^{\CFT}_{\alpha'}}$ in \eqref{eq:numerics} is given purely in terms of CFT quantities for a proper choice of CFT and boundary conditions. It is therefore universal. In contrast, the finite size correction $\mathcal{F}^n_{\alpha \alpha'}$ depends on the microscopic details of the model. To confirm this, we repeated the above calculation for the same field strength $\lambda=1$ but other values of the anisotropy $\gamma \in (0,1]$, that is, for other critical quantum spin chains in the same universality class, and found indeed the same Ising CFT coefficients $\bra{v^{\CFT}_{\alpha}} H^{\CFT}_n \ket{v^{\CFT}_{\alpha'}}$. On the other hand, for $0\leq\lambda<1$ and $\gamma = 0$, which corresponds to the universality class of the free boson CFT, we recovered once again the structure of \eqref{eq:numerics}, but this time the coefficients $\bra{v^{\CFT}_{\alpha}} H^{\CFT}_n \ket{v^{\CFT}_{\alpha'}}$ were dictated by the boson CFT (see Appendices II and V for details).

\textit{Discussion.---} Bipartite entanglement, in terms of both the entanglement entropy $S_A$ and of the entanglement spectrum $\{E_{\alpha}\}$, has been long known to capture universal properties of a quantum phase transition (e.g. in critical quantum spin chains) and of non-trivial gapped phases of matter (e.g. in chiral topological order). In this work, going beyond the bipartite entanglement paradigm, we have argued and numerically demonstrated that not just the eigenvalues $\{E_{\alpha}\}$ of the entanglement Hamiltonian $K_A$, but also its eigenvectors $\{\ket{v_{\alpha}}\}$ (or Schmidt vectors), display emergent universal behaviour in the way they relate to the original lattice Hamiltonian density $h_{j,j+1}$, as investigated in terms of the lattice operators $H_n$. Our main result, Conjecture 2 (i.e. Eq. \eqref{eq:conjecture2}), can be interpreted as probing how the entanglement Hamiltonian $K_A$ is universally embedded in the factorization \eqref{eq:factorization} of the Hilbert space $\mathbb{V}^{A}$. Note that a salient property of $h_{j,j+1}$ is its locality with respect to that factorization. Off criticality, $h_{j,j+1}$ is still local, but no longer known to be related to $K_A$ in any simple way.

We can further refine our construction by directly building a lattice version of the entanglement Virasoro generators $L_n$ on region $A$, and not just the linear combinations $H_n = \f{1}{2}(L_n + L_{-n})$. Here $L_n$ is a weighted sum of both the lattice Hamiltonian density $h_{j,j+1}$ and the lattice momentum density $p_{j,j+1,j+2}$ (obtained from $h_{j,j+1}$, see Appendix IV). Overall, these results imply the surprising ability to manipulate, using simple local lattice operators such as $H_n$ (or $L_n$), the Schmidt decomposition of the otherwise intricate ground state of a many-body Hamiltonian, in a controlled, well-understood manner. 

Eq. \eqref{eq:conjecture2} and its generalization for $L_n$ are ultimately rooted in the Koo-Saleur formula \cite{koo1994representations}, that equates the lattice Hamiltonian and momentum densities $h_{j,j+1}$ and $p_{j,j+1,j+2}$ with their CFT counterpart $h(x)$ and $p(x)$. So far, the Koo-Saleur formula had found applications in the context of probing the structure of low energy eigenstates of critical quantum spin chains \cite{milsted2017extraction}, that is, in relation to the true Hamiltonian $H$. In this work we have extended its use to exploring the low-energy structure of the entanglement Hamiltonian $K_A$. Natural generalizations of our work include exploring how to (i) connect different conformal towers in the entanglement spectrum, (ii) make missing conformal towers appear (e.g. the conformal tower of the spin $\sigma$ operator in the case of the Ising BCFT), or (iii) build a representation of the Kac-Moody algebra in the presence of an extended symmetry. 

\textit{Acknowledgements.}
The authors are grateful to Yijian Zou for helpful discussions. The authors acknowledge support from Compute Canada.  Research at Perimeter Institute is supported in part by the Government of Canada through the Department of Innovation, Science and Economic Development Canada and by the Province of Ontario through the Ministry of Colleges and Universities.  GV is a CIFAR fellow in the Quantum Information Science Program. X is formerly known as Google[x] and is part of the Alphabet family of companies, which includes Google, Verily, Waymo, and others (www.x.company).

\bibliography{reference}

\newpage

\section{Appendix I: Entanglement Virasoro algebra in a CFT}\label{sect1}

In this Appendix we derive the entanglement Virasoro algebra for a finite interval $A$, for the ground state of a CFT on the real line (that is, in either 1+1 spacetime dimensions or 2-dimensional Euclidean space). To simplify the notation, in this appendix we do not use the $\CFT$ superscript used in the main text to distinguish CFT objects from those on the lattice.

First we review the path integral representation of the reduced density matrix $\rho_A$ on a finite interval $A$. Then, using conformal transformations, we map that path integral into one on some subregion (a semi-annulus) of the upper half plane. This transformation has been previously used to obtain an expression for the entanglement Hamiltonian $K_A$ in terms of the Hamiltonian density $h(x)$. Here we use it to obtain an expression for the Virasoro generators $L_n$ in terms of the Hamiltonian and momentum densities $h(x)$ and $p(x)$.

\subsection{Path integral representation of the reduced density matrix on a finite interval $A$} 

\begin{figure}
\centering
\subfigure[]{
    \label{fig:Psi_a}
    \includegraphics[width=0.48\linewidth]{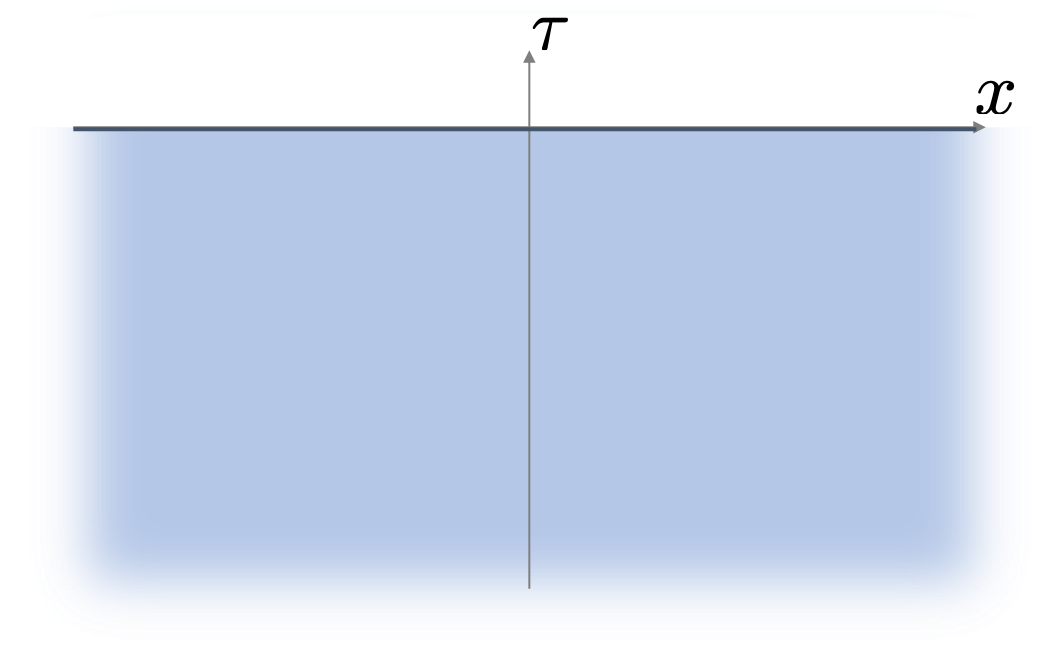}}
\subfigure[]{
    \label{fig:Psi_b}
    \includegraphics[width=0.48\linewidth]{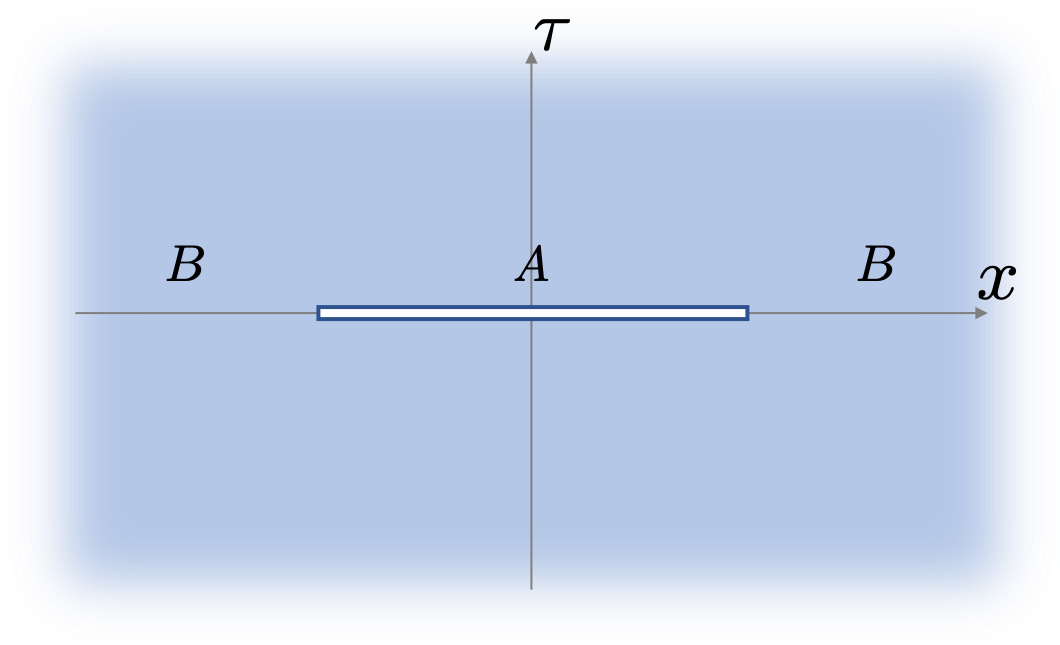}}
\caption{(a) Path integral representation of the ground state $|0\ra$ of a 1+1-dimensional QFT. (b) The reduced density matrix $\rho_A$ on a finite interval $A$. The field $\phi(\tau,x)$ takes values $\Phi_A^-(x^-)$ and $\Phi_A^+(x^+)$ on the lower and upper edges of the open cut.}
\end{figure}

We start by considering a general quantum theory of a field $\phi(\tau,x)$ in 2-dimensional Euclidean space $(\tau,x)$, with action functional $S[\phi]$. Its ground state $|0\ra$ can be prepared, up to normalization, by Euclidean time evolution from $\tau=-\infty$ to $\tau=0$, and therefore it can be represented by a Euclidean path integral over the lower half plane
\beq
\la \Phi(x)|0\ra=\int_{\phi(0,x)=\Phi(x)}[D\phi(\tau<0,x)]e^{-S[\phi]} ,
\eeq
where $|\Phi(x)\ra$ is a field eigenstate with spatial field configuration $\Phi(x)$. This is illustrated in Fig. \ref{fig:Psi_a}. Similarly, a Euclidean path integral over the upper half plane prepares the Hermitian conjugate of the ground state:
\beq
\la 0|\Phi(x)\ra=\int_{\phi(0,x)=\Phi(x)}[D\phi(\tau>0,x)]e^{-S[\phi]}.
\eeq
To compute the reduced density matrix on a finite interval $A\subset \mathbb R$, we trace out the fields on $B$, the complement of $A$ (as represented pictorially by Fig. \ref{fig:Psi_b}):
\beqa
\blgn
&\la\Phi_A^-(x^-)|\rho_A|\Phi_A^+(x^+)\ra=\int [D\Phi_{B}] \la\Phi_A^-\Phi_{B}| 0\ra\la 0  |\Phi_A^+ \Phi_{B}\ra\\
&=\int_{\phi(0^-,x^-\in A)=\Phi_A^-(x^-)}^{\phi(0^+,x^+\in A)=\Phi_A^+(x^+)}[D\phi(\tau,x)]e^{-S[\phi]}.
\elgn
\eeqa

\begin{figure}
\includegraphics[width=2.8in]{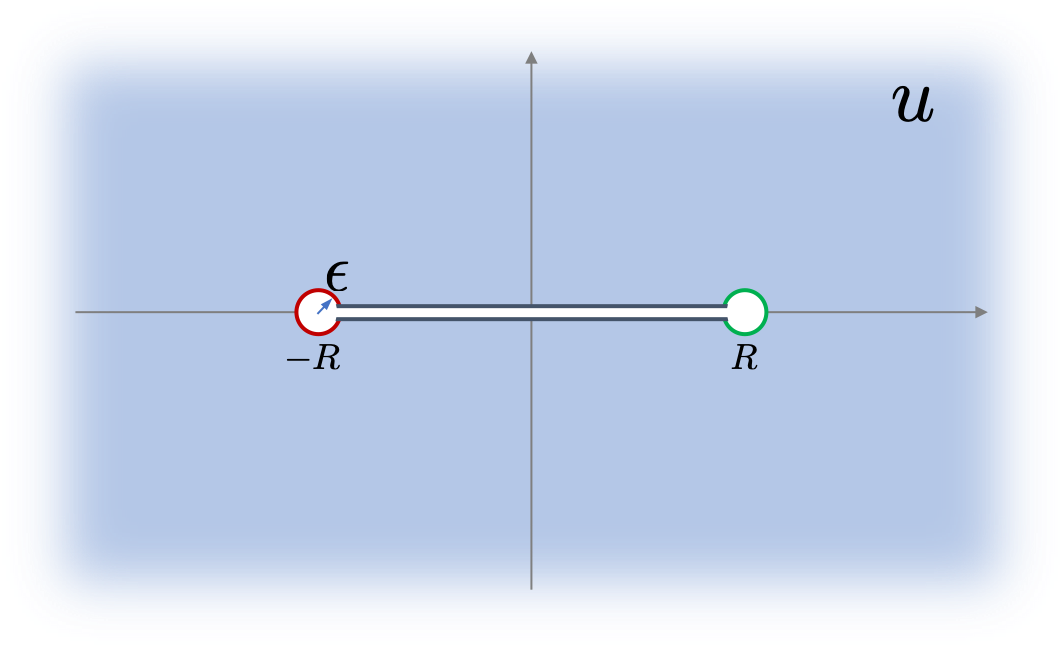}
\caption{The regularized path integral representation of the reduced density matrix $\rho_A$ on a finite interval $A=(-R,R)$ for a 1+1-dimensional CFT. Some conformal boundary conditions are imposed on the red and blue circles.}
\label{fig:rdm_eps}
\end{figure}

The entanglement entropy of $A$, which we regard as a proxy for other objects of interest, will typically be UV divergent. This is due to the presence of quantum entanglement at arbitrarily small distances at the boundaries between $A$ and $B$. In order to regularize it, we follow \cite{cardy2016entanglement} and remove small disks of radius $\ep$ around the boundaries from the path integral, as displayed in Fig. \ref{fig:rdm_eps}. Here $\ep$ plays the role of the UV regulator, analogously to the lattice constant in a lattice regularization (however, later on we will see that on the lattice, the lattice spacing $a$ and $\ep$ differ). We assume $\ep$ is much smaller than the size of the interval. Note that the boundary conditions on the boundaries of the removed disks are yet to be specified for a complete regularization prescription.

\subsection{Entanglement Hamiltonian and entanglement Virasoro algebra on a finite interval $A$}

From now on we restrict our attention to a conformal field theory (CFT). We next show how to define a representation of the Virasoro algebra $L_n$ acting on the Schmidt vectors of the interval $A$ (that is, on the eigenvectors $\ket{v_{\alpha}}$ of $\rho_A$). For $n=0$, the generator $L_0$ will be seen to be proportional to the entanglement Hamiltonian $K_A$.

In the interval $A=(-R,R)$ of Fig. \eqref{fig:rdm_eps}, we must first choose conformal boundary conditions of type $\alpha$ and $\beta$ for the two disks of radius $\ep$ ($\ep \ll R$) around the two boundary points $-R$ and $R$. Conformal boundary conditions, defined to be stable fixed points of the flows of the boundary renormalization group \cite{cardy2016entanglement}, are a natural choice of boundary conditions in a CFT. They determine a universal correction to the entanglement entropy, which is the celebrated Affleck-Ludwig boundary entropy \cite{affleck1991universal}. More generally, they determine the specific boundary CFT (that is, the specific choice of chiral conformal towers of the original CFT) that appear in the entanglement spectrum of the interval $A$.

\begin{figure}
\centering
\subfigure[]{
    \label{fig:rdm_wz_a}
    \includegraphics[width=.48\linewidth]{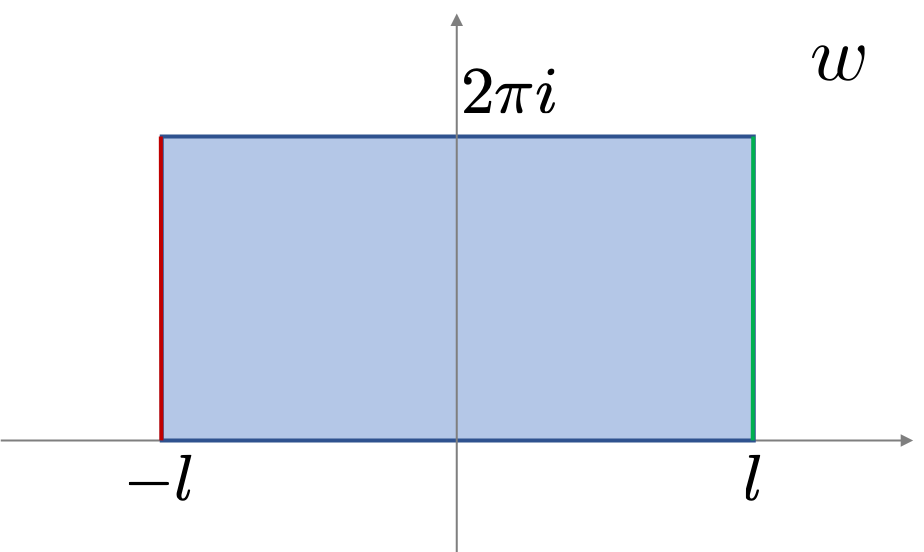}}
\subfigure[]{
    \label{fig:rdm_wz_b}
    \includegraphics[width=.48\linewidth]{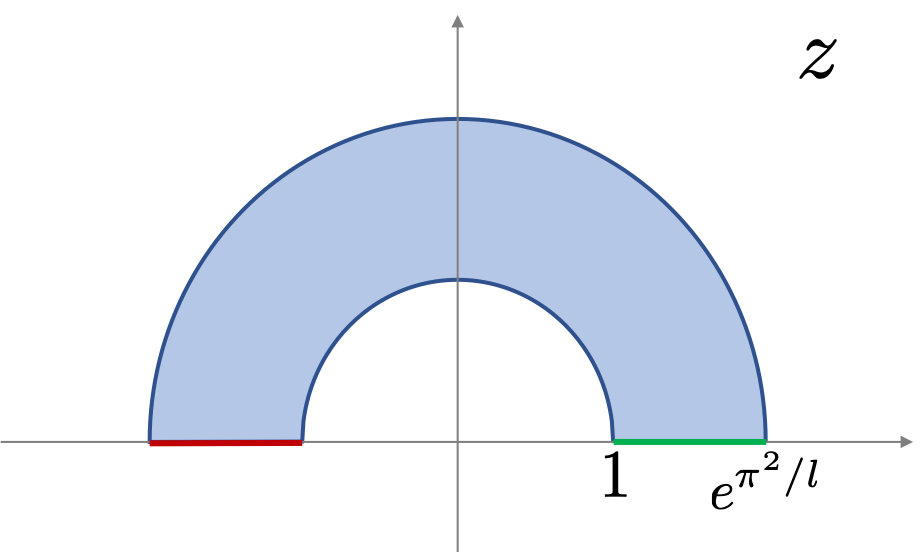}}
\caption{(a) The representation of the reduced density matrix after a conformal transformation $w=\log\left(\f{R+u}{R-u}\right)$. (b) The representation of the reduced density matrix after a further conformal transformation $z=i\exp(-\f{i\pi w}{2l})$.}
\end{figure}

To exploit the power of the conformal symmetry in 2 dimensions, we use complex coordinates $u,\bar u=x\pm i\tau$. Following \cite{cardy2016entanglement}, we apply the conformal transformation 
\beq
w=\log\left(\f{R+u}{R-u}\right), 
\label{eq:wu}
\eeq
to map the region of the path integral in Fig. \ref{fig:rdm_eps} to a rectangle, with height $2\pi$ and width $2l$, where $l=\log\left(\f{2R}{\ep}\right)$, as shown in Fig. \ref{fig:rdm_wz_a}. The original path integral is then map onto a path integral on the strip $-l<\real{w}<l$, with time flowing in the vertical direction. Thus the reduced density matrix $\rho_A$ can be viewed as a thermal state of a Hamiltonian $H_{\al\be}$ on the region $(-l,l)$ with conformal boundary conditions $\alpha$ and $\beta$.
\par Next, we further apply the exponential map
\beq
z=i\exp\left(-\f{i\pi w}{2l} \right),
\label{eq:zw}
\eeq
so that the BCFT is defined on the upper half-plane of $z$. The boundary conditions $\al$ and $\be$ are imposed on the negative and positive real axis respectively. The origina path integral for $\rho_A$ is now defined on a semiannulus comprised between the semicircumferences $|z|=1$ and $|z|=e^{\pi^2/l}$ (see Fig. \ref{fig:rdm_wz_b}). Therefore the reduced density matrix $\rho_A$ is generated by the dilation operator and the entanglement spectrum simply corresponds to the conformal dimensions of the boundary CFT.

Recall that on the upper half-plane, conformal invariance is restricted to conformal transformations that leave the boundary (the real axis) invariant. Because of this constraint, the holomorphic and antiholomorphic transformations are no longer independent, and the full symmetry algebra of the BCFT is given by a single copy of the Virasoro algebra (instead of two copies as in the original CFT). The Virasoro generators $L_n$ in the $z$ coordinates are given by
\beq
L_n=\f{1}{2\pi i}\int_{\mathcal C}\left[ dz~ z^{n+1}T(z)-d\zb~ \zb^{n+1} \Tb(\zb)  \right],
\eeq
where the integration contour $\mathcal C$ is a semicircle going counterclockwise around the origin.

\begin{figure}[]
\includegraphics[width=2in]{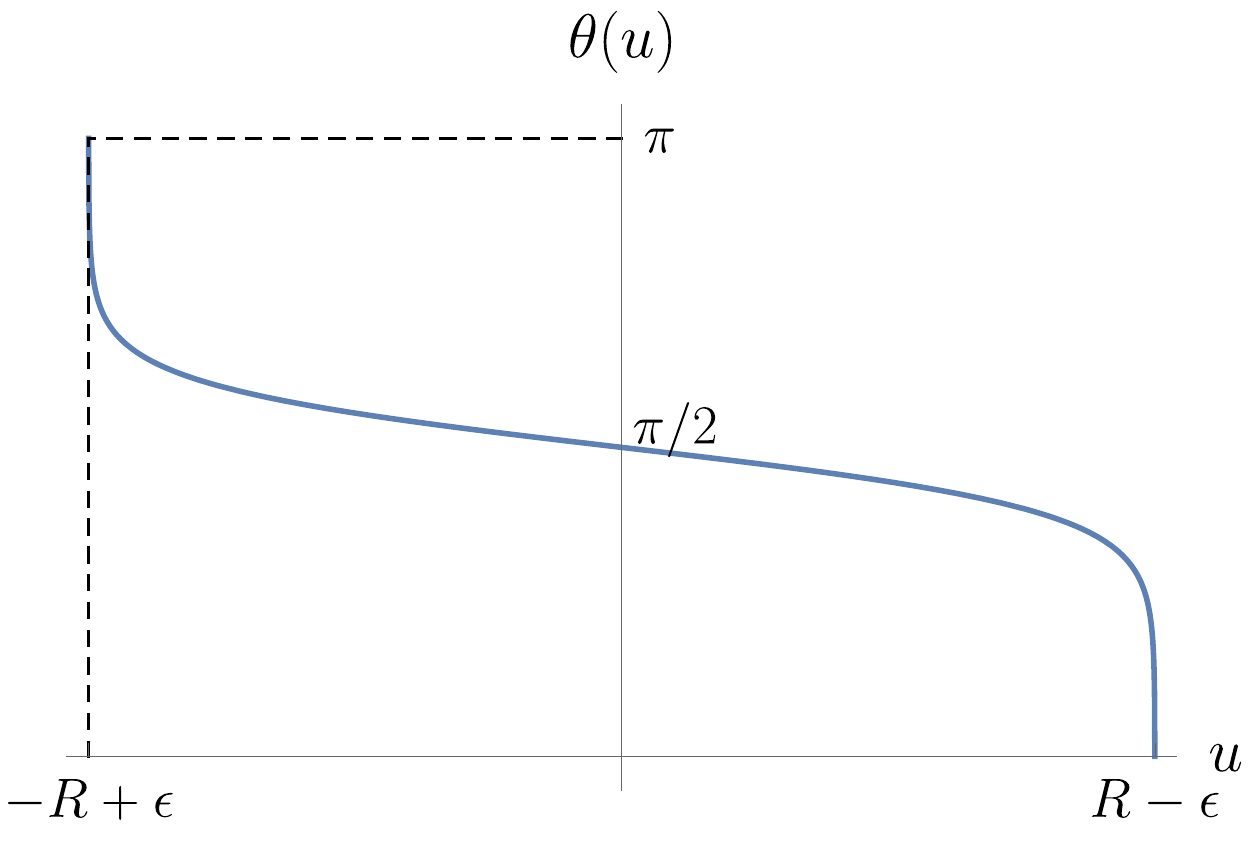}
\caption{Profile $\theta(u)$ for real $u$ and $\ep=10^{-4}R$.}
\label{fig:theta}
\end{figure}

Applying the aforementioned conformal transformation \eqref{eq:zw}, we can transform $L_n$ to the $w$ coordinates:
\beqa
\blgn
	L_n=-\frac{l}{\pi^2}\int_{-l}^l{dw\left[e^{-\frac{in\pi}{2l}(w-l)}T(w)\right.}&\\
	\left.+e^{\frac{in\pi}{2l}(w-l)}\Tb(w)\right]&+\dfrac{c}{24}\delta_{n,0}.
\elgn
\eeqa
We then apply \eqref{eq:wu} to transform it to the $u$ coordinates:
\beqa
\blgn
L_n =&\dfrac{c}{24}\left(1+\frac{4l^2}{\pi^2}\right)\delta_{n,0}\\
-&\dfrac{l}{\pi^2}\int_{-R+\epsilon}^{R-\epsilon}{du}\frac{R^2-u^2}{2R}\left\{e^{in\theta(u)}T(u)+e^{-in\theta(u)}\Tb(u)\right\}.
\elgn
\eeqa
Here the integration contour is along the real axis, from $-R+\ep$ to $R-\ep$, and the function $\theta(u)$ is given by
\begin{equation}
    \theta(u)=\frac{\pi}{2}-\frac{\pi}{2l}\log\left(\frac{R+u}{R-u} \right),
\end{equation}
whose profile for real $u$ is shown in Fig. \ref{fig:theta}.
The expressions for the Virasoro generators can be rewritten in terms of the Hamiltonian and momentum densities:
\beqa
\blgn
h(x) &= -\frac{1}{2\pi}\left( T(x)+\Tb(x)\right)  \\
p(x) &= \frac{1}{2\pi}\left( \Tb(x)-T(x)\right) 
\elgn
\eeqa
	resulting in
\beq
L_0 = \dfrac{l}{\pi }\int_{-R+\epsilon}^{R-\epsilon}{dx\,\dfrac{R^2-x^2}{R}h(x)}+\dfrac{c}{24}+\frac{cl^2}{6\pi^2},
\label{L_0}
\eeq
and
\beq
\blgn
L_n =\dfrac{l}{\pi }\int_{-R+\epsilon}^{R-\epsilon}{dx\,\dfrac{R^2-x^2}{R} }\left[\cos(n\theta(x))h(x)\right.&\\
\left.+i\sin(n\theta(x)) p(x)\right]&
\elgn
\label{L_n}
\eeq
for $n\neq 0$. 

\par The entanglement Hamiltonian $K_A$ of the interval $A$ is given by \cite{casini2011towards,hislop1982modular}
\beq
K_A=\int_{-R+\epsilon}^{R-\epsilon}dx\f{(R^2-x^2)}{2R}h(x)+\text{const}.
\eeq
Comparing it with the expressions of the Virasoro generators, we find that $K_A$ is proportional to $L_0$ up to the addition of a constant:
\beq
K_A=\f{\pi}{2l}L_0+\text{const}.
\label{eq:affine}
\eeq

Accordingly, their eigenvalues have the relation
\beq
E_{\al} = \f{\pi}{2l}h_\al +\text{const}.
\label{eq:spectrum}
\eeq
Since the Schmidt coefficients $\lambda_{\alpha}$ and the entanglement energies $E_{\alpha}$ are related by $\lambda_{\alpha}^2 = e^{-2\pi E_{\alpha}}$, this means that the large-weight Schmidt vectors $\ket{v_\al^A}$ correspond to the low-lying energy eigenstates (or scaling operators with small scaling dimension) of the BCFT. Therefore, the Schmidt vectors constitute a natural representation basis for the Virasoro algebra, whose generators are represented by Eqs. \eqref{L_0}-\eqref{L_n}. In this work we refer to such representation of the Virasoro algebra as the entanglement Virasoro algebra for the interval $A$.

\section{Appendix II: Entanglement Virasoro algebra in Ising and free Boson CFT}
\label{app:operator_content}

In this Appendix we review the operator content compatible with specific boundary conditions for the Ising CFT and the free boson CFT. We also compute the matrix elements of the Virasoro generators in the basis of low lying states. (In this Appendix we drop the $\CFT$ superindex for brevity.)

\subsection{Ising CFT}

The Ising model is described by the Ising CFT at low energy. It has central charge $c=1/2$. The operator content of the Ising CFT with Neumann boundary conditions is shown in Fig. \ref{fig:Ising_towers}. 

The primary fields are the identity $\mathbb{1}$ and energy density $\varepsilon$. The descendant fields can be generated by applying $L_{-n}$ $(n>0)$ to the primary fields. In the following we list the 7 lowest fields ordered by their conformal dimensions (the coefficients are chosen such that the fields are normalized):
\begin{equation*}
\begin{aligned}
&v_1=\mathbb 1,        &&h_1=0; \\
&v_2=\varepsilon,     &&h_2= 1/2; \\
&v_3 =L_{-1}\varepsilon,     &&h_3=3/2;\\
&v_4=2 L_{-2}\mathbb 1,     &&h_4=2;\\
&v_5 =\f 1 2 L_{-1}L_{-1}\varepsilon,     &&h_5=5/2;\\
&v_6 =L_{-1}L_{-2}\mathbb 1,     &&h_6=3;\\
&v_7 =\f 1 6 L_{-1}L_{-1}L_{-1}\varepsilon,    &&h_7=7/2.
\end{aligned}
\end{equation*}
By applying the commutation relations of the Virasoro algebra (Eq. \eqref{eq:VirasoroAlgebra}), we can readily compute the matrix elements of the Virasoro generators $\bra{v_i} L_n\ket{v_j}$ for $i,j=1,\cdots,7$. $L_0$ is simply a diagonal matrix with diagonal elements being the conformal dimensions. $L_1$ and $L_{-1}$ have matrix form 
\begin{equation} \label{eq:IsingCFT}
\left(\begin{array}{ccccccc}
0 & 0 & 0 & 0 & 0 & 0 & 0 \\
0 & 0 & 1 & 0 & 0 & 0 & 0 \\
0 & 0 & 0 & 0 & 2 & 0 & 0 \\          
0 & 0 & 0 & 0 & 0 & 2 & 0\\
0 & 0 & 0 & 0 & 0 & 0 & 3\\
0 & 0 & 0 & 0 & 0 & 0 & 0\\
0 & 0 & 0 & 0 & 0 & 0 & 0
\end{array} \right), ~~~ \left(\begin{array}{ccccccc}
0 & 0 & 0 & 0 & 0 & 0 & 0 \\
0 & 0 & 0 & 0 & 0 & 0 & 0 \\
0 & 1 & 0 & 0 & 0 & 0 & 0 \\          
0 & 0 & 0 & 0 & 0 & 0 & 0\\
0 & 0 & 2 & 0 & 0 & 0 & 0\\
0 & 0 & 0 & 2 & 0 & 0 & 0\\
0 & 0 & 0 & 0 & 3 & 0 & 0
\end{array} \right).
\end{equation}
$L_2$ and $L_{-2}$ have matrix form 
\begin{equation} \label{eq:IsingCFT}
\left(\begin{array}{ccccccc}
0 & 0 & 0 & \frac{1}{2} & 0 & 0 & 0 \\
0 & 0 & 0 & 0 & \frac{3}{2} & 0 & 0 \\
0 & 0 & 0 & 0 & 0 & 0 & \frac{5}{2} \\          
0 & 0 & 0 & 0 & 0 & 0 & 0\\
0 & 0 & 0 & 0 & 0 & 0 & 0\\
0 & 0 & 0 & 0 & 0 & 0 & 0\\
0 & 0 & 0 & 0 & 0 & 0 & 0
\end{array} \right), ~~~ \left(\begin{array}{ccccccc}
0 & 0 & 0 & 0 & 0 & 0 & 0 \\
0 & 0 & 0 & 0 & 0 & 0 & 0 \\
0 & 0 & 0 & 0 & 0 & 0 & 0 \\          
\frac{1}{2} & 0 & 0 & 0 & 0 & 0 & 0\\
0 & \frac{3}{2} & 0 & 0 & 0 & 0 & 0\\
0 & 0 & 0 & 0 & 0 & 0 & 0\\
0 & 0 & \frac{5}{2} & 0 & 0 & 0 & 0
\end{array} \right).
\end{equation}

\subsection{Free boson CFT}

\begin{figure}[h]
	\centering
	\includegraphics[width=0.8\linewidth]{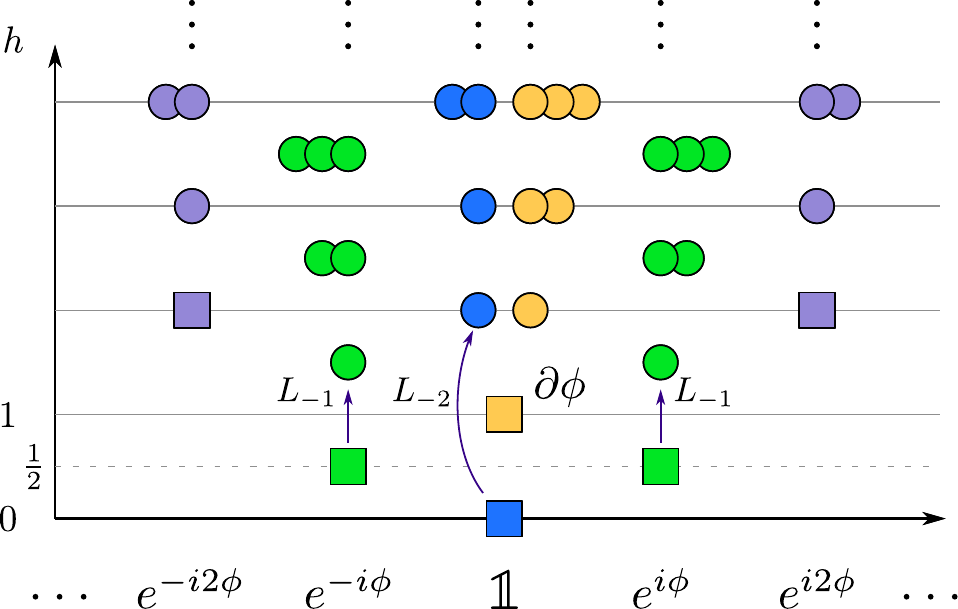}
	\caption{Conformal towers of the chiral boson CFT of compactification radius $r=1$ with Dirichlet boundary conditions.
	}
	\label{fig:boson_towers}
\end{figure}

The XX model (for even number of sites) is described by the free boson CFT of compactification radius $r=1$ at low energy, which has central charge $c=1$. It can also be mapped to two copies of Ising model. The operator content of the boson CFT ($r=1$) with Dirichlet boundary conditions (where the fields are pinned to the same values on the two boundaries) \cite{roy2020entanglement} is shown in Fig. \ref{fig:boson_towers}.

The primary fields are the identity $\mathbb{1}$, $\pa\phi$, and vertex operators $e^{in\phi}$ ($n\in\mathbb Z$). Similarly, we list the 7 lowest fields:
 \begin{equation*}
 \begin{aligned}
 &v_1=\mathbb 1,        &&h_1=0; \\
 &v_2=e^{i\phi},     &&h_2= 1/2; \\
 &v_3 =e^{-i\phi},     &&h_3=1/2;\\
 &v_4=i\pa\phi,     &&h_4=1;\\
 &v_5 =L_{-1}e^{i\phi},     &&h_5=3/2;\\
 &v_6 =L_{-1}e^{-i\phi},     &&h_6=3/2;\\
 &v_7 =\sqrt 2 L_{-2}\mathbb 1,    &&h_7=2.
 \end{aligned}
 \end{equation*}
Again, in this basis, $L_0$ is simply a diagonal matrix consisting of the conformal dimensions. The matrices of $L_1$ and $L_{-1}$ are given by
\begin{equation} \label{eq:IsingCFT}
\left(\begin{array}{ccccccc}
0 & 0 & 0 & 0 & 0 & 0 & 0 \\
0 & 0 & 0 & 0 & 1 & 0 & 0 \\
0 & 0 & 0 & 0 & 0 & 1 & 0 \\          
0 & 0 & 0 & 0 & 0 & 0 & 0\\
0 & 0 & 0 & 0 & 0 & 0 & 0\\
0 & 0 & 0 & 0 & 0 & 0 & 0\\
0 & 0 & 0 & 0 & 0 & 0 & 0
\end{array} \right), ~~~ \left(\begin{array}{ccccccc}
0 & 0 & 0 & 0 & 0 & 0 & 0 \\
0 & 0 & 0 & 0 & 0 & 0 & 0 \\
0 & 0 & 0 & 0 & 0 & 0 & 0 \\          
0 & 0 & 0 & 0 & 0 & 0 & 0\\
0 & 1 & 0 & 0 & 0 & 0 & 0\\
0 & 0 & 1 & 0 & 0 & 0 & 0\\
0 & 0 & 0 & 0 & 0 & 0 & 0
\end{array} \right).
\end{equation}
The matrices of $L_2$ and $L_{-2}$ are given by
\begin{equation} \label{eq:IsingCFT}
\left(\begin{array}{ccccccc}
0 & 0 & 0 & 0 & 0 & 0 & \f{1}{\sqrt{2}} \\
0 & 0 & 0 & 0 & 0 & 0 & 0 \\
0 & 0 & 0 & 0 & 0 & 0 & 0 \\          
0 & 0 & 0 & 0 & 0 & 0 & 0\\
0 & 0 & 0 & 0 & 0 & 0 & 0\\
0 & 0 & 0 & 0 & 0 & 0 & 0\\
0 & 0 & 0 & 0 & 0 & 0 & 0
\end{array} \right), ~~~ \left(\begin{array}{ccccccc}
0 & 0 & 0 & 0 & 0 & 0 & 0 \\
0 & 0 & 0 & 0 & 0 & 0 & 0 \\
0 & 0 & 0 & 0 & 0 & 0 & 0 \\          
0 & 0 & 0 & 0 & 0 & 0 & 0\\
0 & 0 & 0 & 0 & 0 & 0 & 0\\
0 & 0 & 0 & 0 & 0 & 0 & 0\\
\f{1}{\sqrt{2}} & 0 & 0 & 0 & 0 & 0 & 0
\end{array} \right).
\end{equation}

\section{Appendix III:  Free fermion solution to the quantum XY model}
\label{app:fermion_solution}

In this Appendix, we study the quantum XY model defined in terms of the following Hamiltonian
\begin{equation}
    H^{XY}
    = - \f 1 2\sum_{j=-\infty}^{\infty} \left(\frac{1+\gamma}{2}\sigma^x_{j}\sigma^x_{j+1} +  \frac{1-\gamma}{2}\sigma^y_{j}\sigma^y_{j+1} + \lambda \sigma^z_j\right).
\end{equation}
This model has been thoroughly studied in \cite{lieb1961two,katsura1962statistical,barouch1971statistical,vidal2003entanglement,latorre2003ground}. In the following we present an exact solution using free fermion representation.

\begin{figure}
    \centering
    \includegraphics[width=0.6\columnwidth]{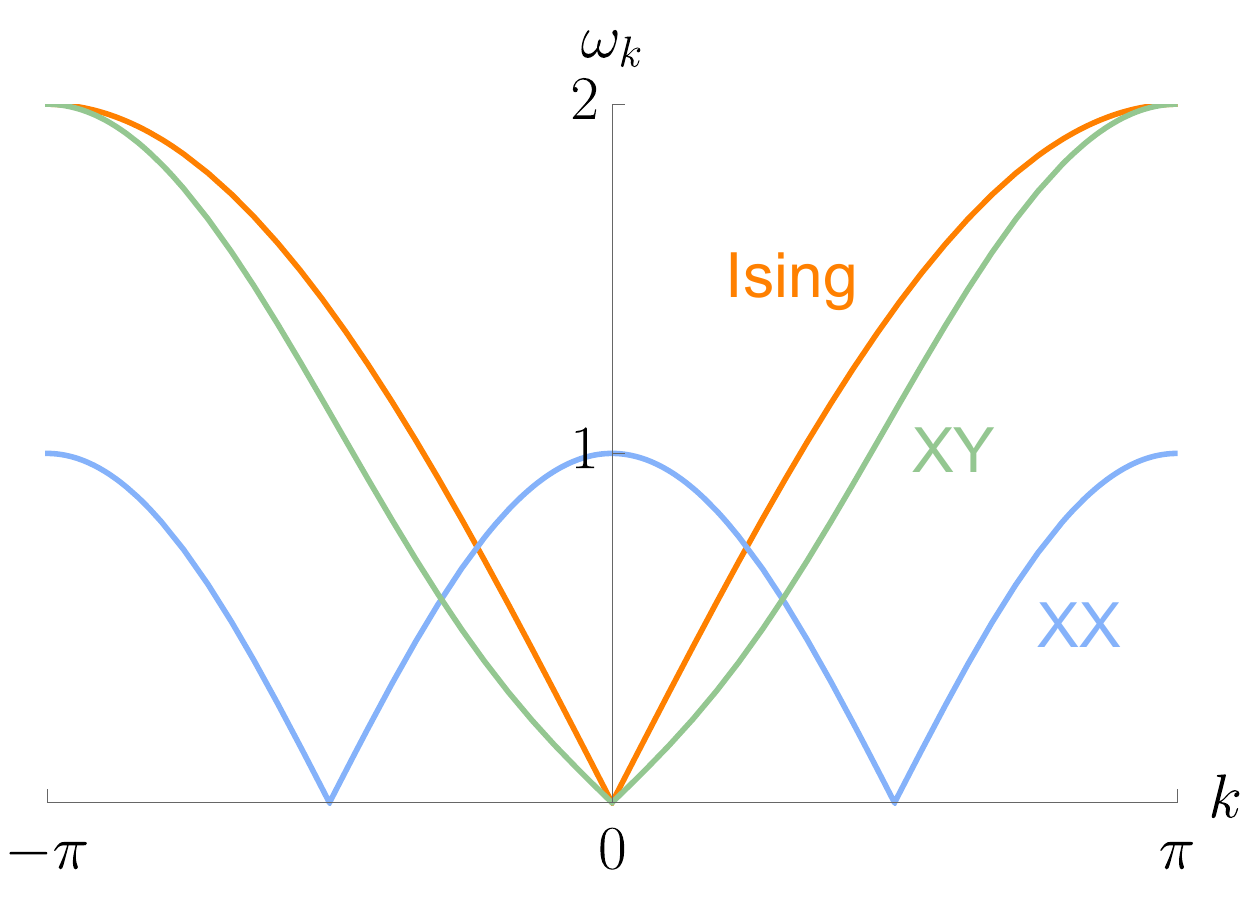}
    \caption{The dispersion relations for the critical Ising model ($\ga=1, \lambda=1$), the critical XX model ($\ga=0, \lambda=0$) and the critical XY model ($\ga=0.9, \lambda=1$).}
    \label{fig:XYenergy}
\end{figure}

This Hamiltonian can be written in terms of free fermionic variables, thus permitting its study via the correlation matrix formalism. The required change of variables is given by the Jordan-Wigner transformation
\begin{equation}
    a_n = \left(\prod_{j<n}{\sigma^z_j}\right)\dfrac{\sigma^x_n-i\sigma_n^y}{2},
\end{equation}
which guarantees that the resulting operators satisfy canonical anticommutation relations
\begin{equation}
    \{a_i^{\phantom{\dagger}},a_j^\dagger\}=\delta_{ij}.
\end{equation}
The resulting Hamiltonian reads
\begin{align}
  H_{XY}=\frac{1}{2} \sum_{j=-\infty}^{\infty}
&\left[ a_{j+1}^\dagger a_{j} + a_j^\dagger a_{j+1}\right.\nonumber\\ 
& \left.+ \gamma \left(a_j^\dagger a_{j+1}^\dagger + a_{j+1} a_j\right)- 2\lambda a_j^{\dagger} a_j \right]+\text{const}.  
\end{align}
This Hamiltonian can be diagonalized in momentum space, so we change variables again
\begin{equation}
    a_k = \dfrac{1}{\sqrt{2\pi}}\sum ^\infty_{j=-\infty}{a_j e^{ijk}},\qquad k \in(-\pi,\pi],
\end{equation}
and obtain
\begin{align}
     H_{XY}=&\int_{-\pi}^\pi{dk}\left[(\cos{k}-\lambda)\, a_k^\dagger a_k^{\phantom{\dagger}}\phantom{\dfrac{1}{1}}\right.\nonumber\\
    &~~~~~~~~~~\,+\left.\dfrac{i\gamma}{2}\sin{k}  \left(a_ka_{-k}+a^\dagger_ka^\dagger_{-k}\right)\right]+\text{const}.
\end{align}
The final step towards diagonalizing this Hamiltonian requires that we perform a Bogoliubov (canonical) transformation to define a new fermionic variable:
\begin{align}
    b_k \equiv \cos{\frac{\theta_k}{2}}a_k-i\sin{\frac{\theta_k}{2}}a_{-k}^\dagger,
\end{align}
where the angular variable $\theta_k$ satisfies
\begin{align}
    \cos\theta_k&=\dfrac{\cos k - \lambda}{\sqrt{(\cos k -\lambda)^2+\gamma^2\sin^2{k}}},\\
    \sin\theta_k&=\dfrac{-\gamma\sin{k}}{\sqrt{(\cos k -\lambda)^2+\gamma^2\sin^2{k}}}.
\end{align}
In terms of $b^{\phantom{\dagger}}_k,b_k^\dagger$, the Hamiltonian is then finally diagonal
\begin{equation}
    H_{XY}=\int_{-\pi}^\pi{dk\;\omega_k b_k^\dagger b^{\phantom{\dagger}}_k}+\text{const}.
\end{equation}
with the dispersion relation given by
\begin{equation}
    \omega_k = \sqrt{(\cos k -\lambda)^2+\gamma^2\sin^2{k}}.
\end{equation}
The dispersion relations for a few special cases are shown in Fig. \ref{fig:XYenergy}. 

\begin{figure}
    \centering
    \includegraphics[width=0.6\columnwidth]{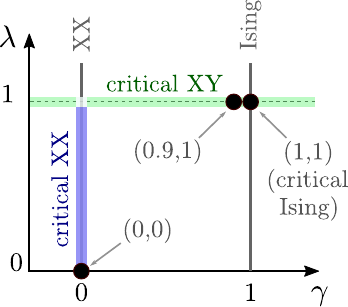}
    \caption{Phase diagram of the quantum XY model. The lines $\gamma=0$ and $\gamma = 1$ correspond to the XX and Ising models, respectively. The critical XY model ($\gamma\neq 0, \lambda = 1$) (including the critical quantum Ising model) belongs to the universality class of the Ising CFT. The critical XX model $(\gamma = 0, 0\leq\lambda<1)$ belongs to the universality class of the free Boson CFT. The point $(\gamma,\lambda) = (0,1)$ corresponds to a scale-invariant but not conformally invariant theory, whose dispersion relation is quadratic. We have marked the three parameter choices that we study in this paper.}
    \label{fig:phase_diag}
\end{figure}

For $\lambda=1$ and $\gamma \neq 0$, the model is critical and falls within the Ising CFT universality class. For $0\leq\lambda<1$ and $\gamma = 0$, the model is also critical, but within the universality class of the free Boson CFT, see Fig. \ref{fig:phase_diag}. Note that for the critical XY model, in order for the CFT derivation to work, we need to rescale the Hamiltonian $H_{XY}$ by a factor of $\f 1 \ga$ so that the velocity of low-energy excitations is 1.

In the ground state all $b_k$ modes are empty, resulting in the correlation functions
\begin{equation}
    \langle b_kb_q\rangle = 0,\qquad \langle b^{\phantom{\dagger}}_kb^\dagger_q\rangle =\delta(k-q).
\end{equation}
The correlation functions in the original fermionic variables can be obtained by using
\begin{equation}
    a_n = \int_{-\pi}^\pi{\dfrac{dk}{\sqrt{2\pi}}}e^{ink}\left(\cos{\frac{\theta_k}{2}}b_k+i\sin{\frac{\theta_k}{2}}b_{-k}^\dagger\right).
\end{equation}
In this way we have
\begin{equation}
	\begin{aligned}
	\langle a^{\phantom{\dagger}}_n a^\dagger_m\rangle &= \int_{-\pi}^{\pi}{\dfrac{dk}{2\pi}e^{i(n-m)k}\cos^2\left(\frac{\theta_k}{2}\right)}\\
	&= \dfrac{\delta_{n,m}}{2}
	+\int_{-\pi}^{\pi}{\dfrac{dk}{4\pi}\dfrac{(\cos k-\lambda) e^{i(n-m)k}}{\sqrt{(\cos k -\lambda)^2+\gamma^2\sin^2{k}}}}.
	\label{corr1}
	\end{aligned}
	\end{equation}
	and
	\begin{equation}
	\begin{aligned}
	\langle a_n a_m\rangle &= \dfrac{-i}{2}\int_{-\pi}^{\pi}{\dfrac{dk}{2\pi}e^{i(n-m)k}\sin\theta_k}\\
	&= \dfrac{i}{2}\int_{-\pi}^{\pi}{\dfrac{dk}{2\pi}\dfrac{\gamma e^{i(n-m)k}\sin{k}}{\sqrt{(\cos k -\lambda)^2+\gamma^2\sin^2{k}}}},
	\label{corr2}
	\end{aligned}
	\end{equation}
The expressions of correlation functions in Eqs.(\ref{corr1}, \ref{corr2}) simplify a lot for special cases. For the Ising model ($\ga=1, \lambda=1$):
	\begin{align}
	\langle a^{\phantom{\dagger}}_n a^\dagger_m\rangle_\text{Ising}&= \dfrac{\delta_{n,m}}{2}+\frac{1}{\pi\left(4(n-m)^2-1\right)},\\
	\langle a_n a_m\rangle_\text{Ising}&= \frac{2(m-n)}{\pi\left(4(n-m)^2-1\right)}.
	\end{align}
	For the XX model ($\ga=0, \lambda=0$):
	\begin{align}
	\langle a^{\phantom{\dagger}}_n a^\dagger_m\rangle_\text{XX}&=\frac 1 2 \operatorname{sinc}\left(\frac{\pi(n-m)}{2}\right), \\
	\langle a_n a_m\rangle_\text{XX}&=0.
	\label{eq:anamXX}
	\end{align}
	Eq.(\eqref{eq:anamXX}) reflects the U(1) symmetry of the XX model.
	
If we define $A$ as the interval including sites $1, \ldots, L$, then Eqs. \eqref{corr1}-\eqref{corr2} allow us to construct the correlation matrix $\Gamma_A$ for $A$ simply by restricting ourselves to fermionic operators operators within $A$:
\begin{equation}
    \Gamma_A \equiv  \left(\begin{array}{cc}
        \langle a_i^{\dagger}a^{\phantom{\dagger}}_j\rangle &  \langle a_i^{\dagger}a_j^{\dagger}\rangle\\[1mm]
        \langle a_ia_j\rangle &  \langle a^{\phantom{\dagger}}_ia_j^{\dagger}\rangle
    \end{array}\right)\qquad i,j=1,\ldots,L
\end{equation}
This correlation matrix can be brought to a diagonal form by means of a canonical transformation on the $a_1,\ldots,a_L$ modes, much like the Bogoliubov transformation we used above:
\begin{align}
    a^{\phantom{\dagger}}_i &\longrightarrow  c^{\phantom{\dagger}}_i = A_{ij}a^{\phantom{\dagger}}_j + B_{ik}a_k^\dagger\\
    a^\dagger_i &\longrightarrow c^{\dagger}_i = B^*_{ik}a^{\phantom{\dagger}}_k + A^*_{ij}a_j^\dagger
\end{align}
For this transformation to be canonical, we have to impose a unitarity condition for the following block matrix:
\begin{equation}
   U= \left(\begin{array}{cc}
    A & B\\
    B^* & A^* 
    \end{array}\right)\implies U^\dagger=U^{-1}
    \label{canonical_transf}
\end{equation}
Thankfully there is an easy method to obtain such matrices $A$ and $B$ by diagonalizing the correlation matrix $\Gamma_A$, which is Hermitian and thus can be diagonalized by means of a unitary transformation. Now, because of its block structure, the eigenvalues of $\Gamma_A$ come in pairs $(\nu, 1-\nu)$. Indeed, if 
\begin{equation}
    \left(\begin{array}{c}
         v \\
         w 
    \end{array}\right)
\end{equation}
where $v,w\in\mathbb C^L$ is an eigenvector of $\Gamma_A$ with eigenvalue $\nu$ then
\begin{equation}
    \left(\begin{array}{c}
         w^*  \\
         v^* 
    \end{array}\right)
\end{equation}
is another eigenvector with eigenvalue $1-\nu$, where $\nu\in[0,1]$ since $\Gamma_A$ is positive semidefinite. Thus, we can choose the unitary $U$ that diagonalizes it to have the form \eqref{canonical_transf}. All this allows us to write $\Gamma_A$ as a direct sum of correlation matrices for a family of uncorrelated fermionic variables $\{c_i^{\phantom{\dagger}}c_i^{\dagger}\}_{i=1}^L$, that is
\begin{equation}
\Gamma_A = \bigoplus_{i=1}^L{\left(\begin{array}{cc}
        \langle c_i^{\dagger}c^{\phantom{\dagger}}_i\rangle &  \langle c_i^{\dagger}c_i^{\dagger}\rangle\\[1mm]
        \langle c_ic_i\rangle &  \langle c^{\phantom{\dagger}}_ic_i^{\dagger}\rangle
    \end{array}\right)}=\bigoplus_{i=1}^L{\left(\begin{array}{cc}
        \nu_i &  0\\[1mm]
        0 &  1-\nu_i
    \end{array}\right)},
\end{equation}
where $0<\nu_i<\frac 1 2$. These uncorrelated fermionic degrees of freedom (which are non-local linear combinations of the original fermion variables), give us the structure of the density matrix $\rho_A$. Indeed, we have
\begin{equation}
    \rho_A = \dfrac{\exp{\left(-\sum_{i=1}^L{e_ic_i^{\dagger}c^{\phantom{\dagger}}_i}\right)}}{\tr\left[{\exp{\left(-\sum_{i=1}^L{e_ic_i^{\dagger}c^{\phantom{\dagger}}_i}\right)}}\right]},
\end{equation}
where we have defined
\begin{equation}
    e_i \equiv \log{\dfrac{1-\nu_i}{\nu_i}}.
\end{equation}
Thus the Schmidt vectors $\ket{v_{\alpha}}$ are of the form
\begin{equation}
\ket{v_{\alpha}}=\left( \prod_{j\in J_\alpha}{c_j^\dagger}\right)\hspace{-1mm}\ket{v_1},\qquad J_\alpha\subseteq\{1,\ldots,2N\},
\end{equation}
where $\ket{v_1}$ is the dominant Schmidt vector satisfying 
\begin{equation}
c_j\ket{v_1} = 0,\qquad j=1,\ldots, 2N.
\end{equation}
The entanglement spectrum $\{E_\alpha\}$, is given by summing up the single-particle spectra$\{e_j\}$:
\begin{equation}
    E_\alpha=\sum_{j\in J_\alpha}e_j+\text{const},
    \label{eq:full_single_spectrum}
\end{equation}
where the constant is determined by the normalization of $\rho_A$.

\section{Appendix IV: Entanglement Virasoro algebra for the XY quantum spin chain}
\label{app:EV_XY}

In this Appendix we derive the entanglement Virasoro algebra for the quantum XY model. 

In order to apply Eqs. \eqref{L_0} and \eqref{L_n} in the context of a critical spin chain, they need to be discretized. A natural way to do it follows from taking the Hamiltonian terms $h_{j,j+1}$ as the \textit{lattice Hamiltonian density} and defining from them the {lattice momentum density} by
\begin{equation}
p_{j-1,j,j+1}\equiv -i[h_{j-1,j}, h_{j,j+1}].
\label{eq:momentum_density}
\end{equation}
This definition for $p_j$ was proposed in \cite{milsted2017extraction}, and is motivated as a lattice version of energy-momentum conservation $\partial_t h(x,t) = \partial_x p(x,t)$, where $\partial_t$ is implemented by the commutator with the Hamiltonian $H$ and $\partial_x$ is replaced with a finite difference on the lattice (with unit lattice spacing),
\begin{equation}
\partial_t h_{j,j+1}= i[H, h_{j,j+1}] = p_{j-1,j,j+1}-p_{j,j+1,j+2}.
\end{equation}

For the quantum XY model, the lattice Hamiltonian density reads
\begin{eqnarray}
	h^{XY}_{j,j+1}	&=& - \f 1 4 \left((1+\gamma)\sigma^x_{j}\sigma^x_{j+1} +  (1-\gamma)\sigma^y_{j}\sigma^y_{j+1}\right) ~~~~~~~\\
	&&~~~+ \frac{1}{4}\left(\lambda \sigma^z_j+ \lambda \sigma^z_{j+1}\right).
\end{eqnarray}
In terms of fermionic variables it reads
\begin{eqnarray}
h^{XY}_{j,j+1} &=&  \f 1 2 \left(a_{j+1}^\dagger a^\phdag_j+\ga a_{j+1}a_j+\text{h.c.}\right)~~~~~~\\
&&~~~~~ - \f \lambda 2 \left(a_{j}^\dagger a^\phdag_j+ a_{j+1}^\dagger a^\phdag_{j+1}\right).
\end{eqnarray}
Using Eq. \eqref{eq:momentum_density} we can find the lattice momentum density
\begin{equation}
\begin{split}
p^{XY}_{j-1,j,j+1}=&\f{i}{4} \left[ (1-\ga^2)a_{j+1}^\da a_{j-1} \right.~~~~~~~~~~~~~~~~~\\
&-\lambda (a_j^\da a_{j-1}+a_{j+1}^\da a_{j}) 
 \vphantom{a_{j}^\da} \\
 &\left.+\ga\lambda (a_j a_{j-1}-a_{j+1} a_{j})  \right] +\text{h.c.}
\end{split}
\end{equation}

In terms of these lattice densities, the integrals from \eqref{L_0}-\eqref{L_n} can be discretized to Riemann sums:
\begin{align}
L_0 =& \dfrac{l}{\pi }\sum_{j=1}^{2N-1}{\dfrac{N^2-x_{j+\frac{1}{2}}^2}{N}h_{j,j+1}}+\dfrac{c}{24}+\frac{cl^2}{6\pi^2},\\
L_n =&\dfrac{l}{\pi }\sum_{j=1}^{2N-1}{\dfrac{N^2-x_{j+\frac{1}{2}}^2}{N} \cos(n\theta(x_{j+\frac{1}{2}}))h_{j,j+1}}\nonumber\\
+&\dfrac{il}{\pi }\sum_{j=1}^{2N-2}{\dfrac{N^2-x_{j+1}^2}{N}\sin(n\theta(x_{j+1}))p_{j,j+1,j+2}},
\label{L_nlat} 
\end{align}
where $x_j\equiv j-N-\frac{1}{2}$ is the discrete position variable (in some cases the lattice densities can be assigned carefully chosen positions to reduce the finite-size effect). 

Notice that this reduces to Eqs. \eqref{eq:H0lat}, \eqref{eq:Hnlat} if we consider linear combinations of $L_n$ and $L_{-n}$
\begin{align}
H_n &= \f 1 2(L_n + L_{-n}) \\
&=\dfrac{l}{\pi }\sum_{j=1}^{2N-1}{\dfrac{N^2-x_{j+\frac{1}{2}}^2}{N} \cos(n\theta(x_{j+\frac{1}{2}}))h_{j,j+1}},
\end{align}
which  only depend on the Hamiltonian densities but not on the momentum densities.

\section{Appendix V: Numerical analysis for the critical XY model and XX model}
\label{app:XY_XX}

In the main text we have demonstrated that the Ising model supports the two conjectures. In the appendix, we show that these two conjectures also hold for the critical XY model and XX model.

\subsection{XY model ($\lambda=1, \ga=0.9$)}
From the entanglement spectrum for interval size $2N = \{64, 128, 256, 512, 1024\}$, we find the approximate value $\ep\approx0.042$. The first few entries of the energy spectrum $\{h_\al\}$ for $2N=1024$ are
\begin{equation}
\{0, 0.50,1.51,2,01,2.57,3.07,3,68,4.08,4.18, \cdots\},
\end{equation}
This matches the lower part of the exact spectrum
\begin{equation}
\{0, 1/2, 3/2, 2, 5/2, 3, 7/2, 4, 4, \cdots\},
\end{equation}
of the Ising chain with Neumann boundary conditions \cite{francesco2012conformal}, or the Cardy state $|\sigma\rangle$, as shown in Fig. \ref{fig:Ising_towers}.

\begin{figure}[h]
	\centering
    \includegraphics[width=0.5\linewidth]{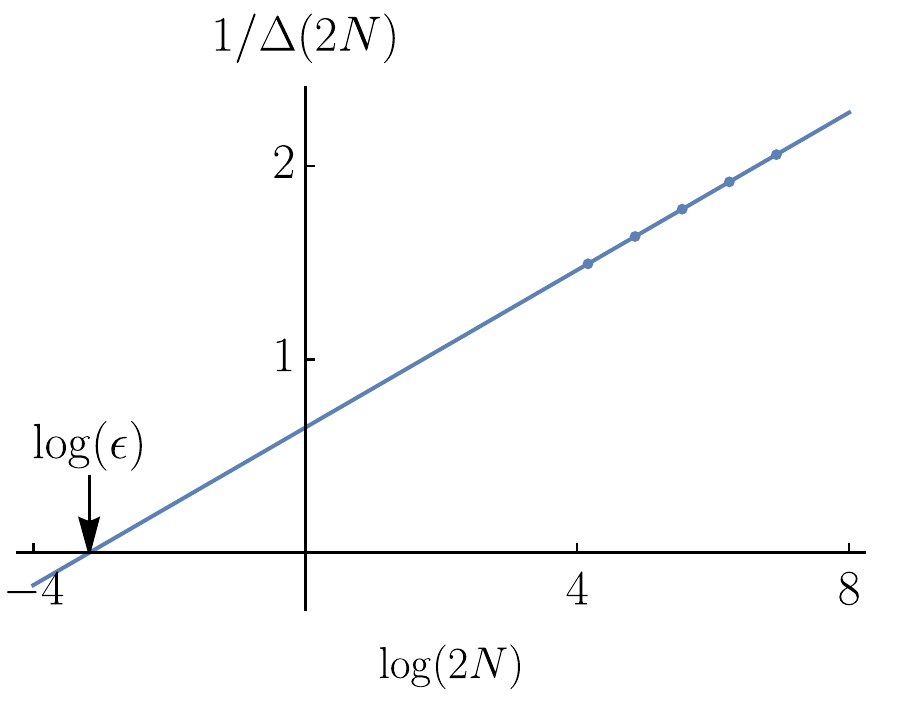}%
	\includegraphics[width=0.5\linewidth]{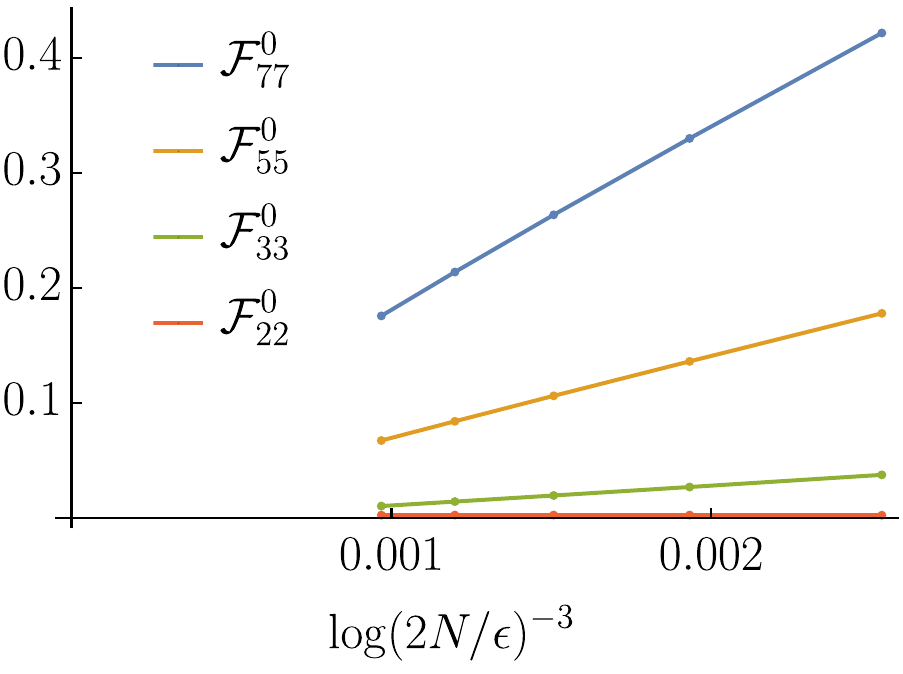}\vspace{4pt}
	\includegraphics[width=0.5\linewidth]{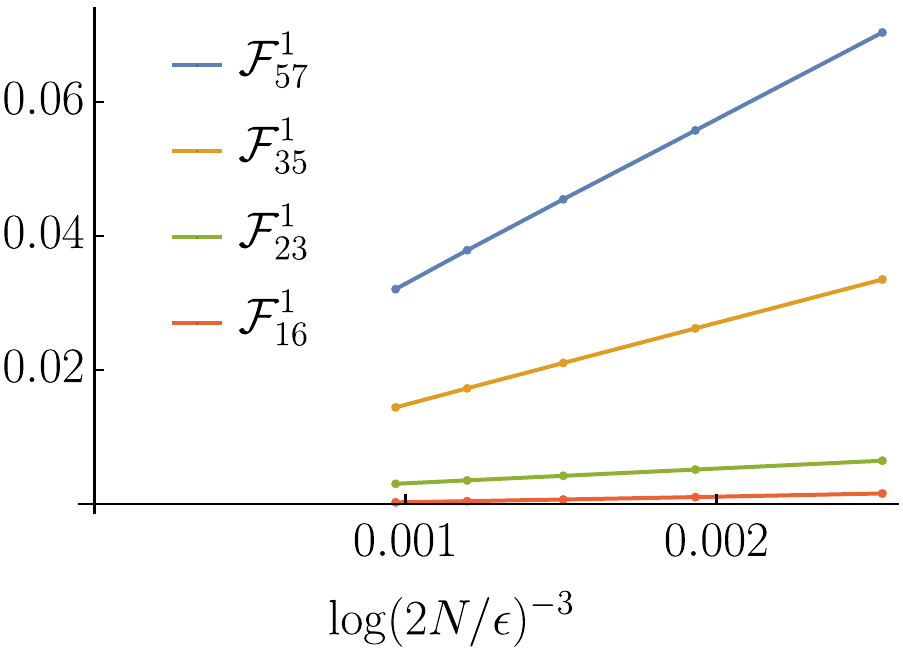}%
	\includegraphics[width=0.5\linewidth]{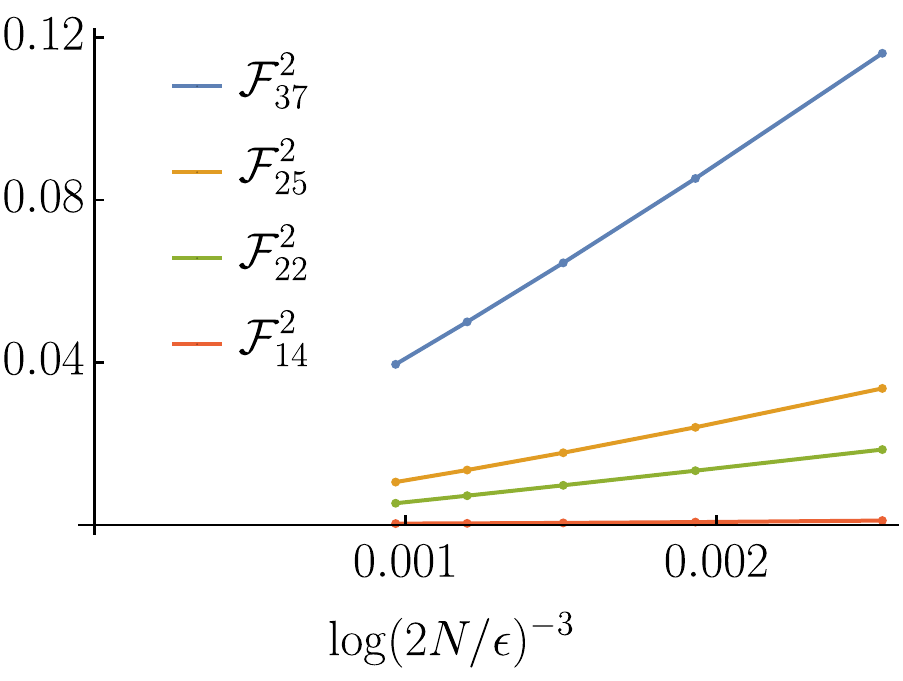}
	\caption{Estimate of $\ep$ by linear extrapolation and finite size corrections $\mathcal{F}^{n}_{\alpha\alpha'}$ for the XY model ($\lambda=1, \ga=0.9$) with interval size $2N = \{64, 128, 256, 512, 1024\}$.
	}
	\label{fig:XY_Hn}
\end{figure}

The matrix elements of $H_n$ are also shown to converge to the exact values, see Fig. \ref{fig:XY_Hn}. 

\subsection{XX model ($\lambda=0, \ga=0$)}
From the entanglement spectrum for interval size $2N =\{512, 1024, 2048, 4096, 8192\}$, we find the approximate value $\ep\approx0.042$. The first few entries of the energy spectrum $\{h_\al\}$ for $2N=8192$ are
\begin{equation}
\{0,0.50,0.50,1.00,1.51,1.51,2.01, 2.01, \cdots\},
\end{equation}
This matches the lower part of the exact spectrum
\begin{equation}
\{0, 1/2, 1/2, 1, 3/2, 3/2, 2, 2 \cdots\},
\end{equation}
of the free boson CFT of compactification $r=1$ with Dirichlet boundary conditions, as shown in Fig. \ref{fig:boson_towers}.

\begin{figure}[h]
	\centering
	\includegraphics[width=0.5\linewidth]{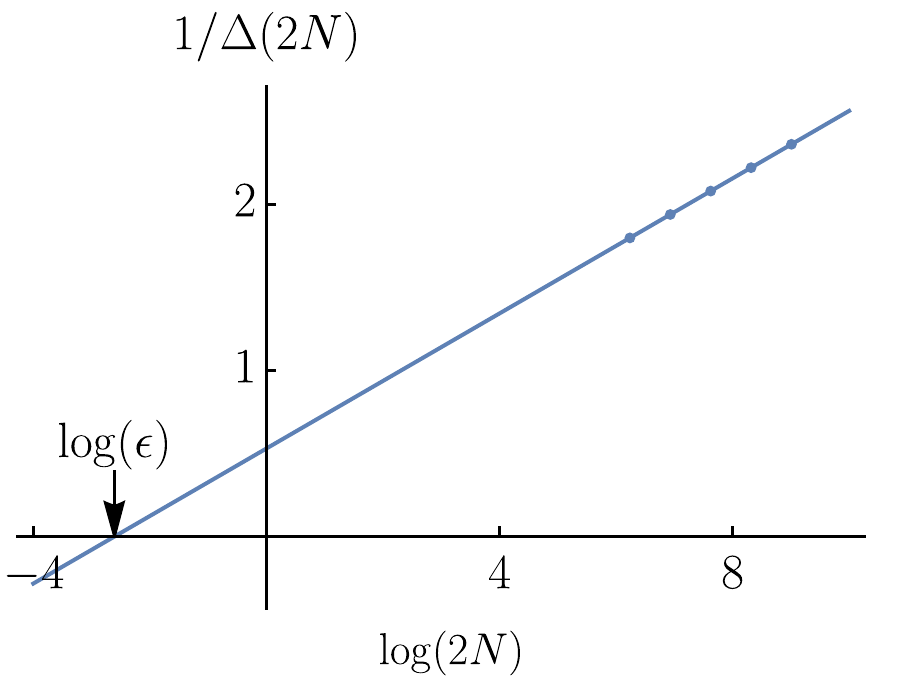}%
	\includegraphics[width=0.5\linewidth]{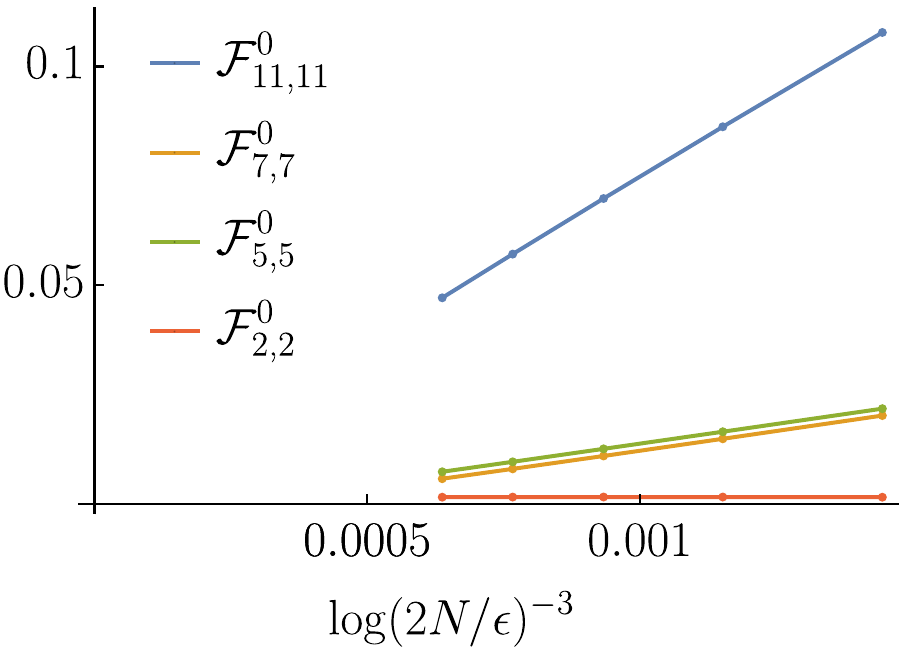}\vspace{4pt}
	\includegraphics[width=0.5\linewidth]{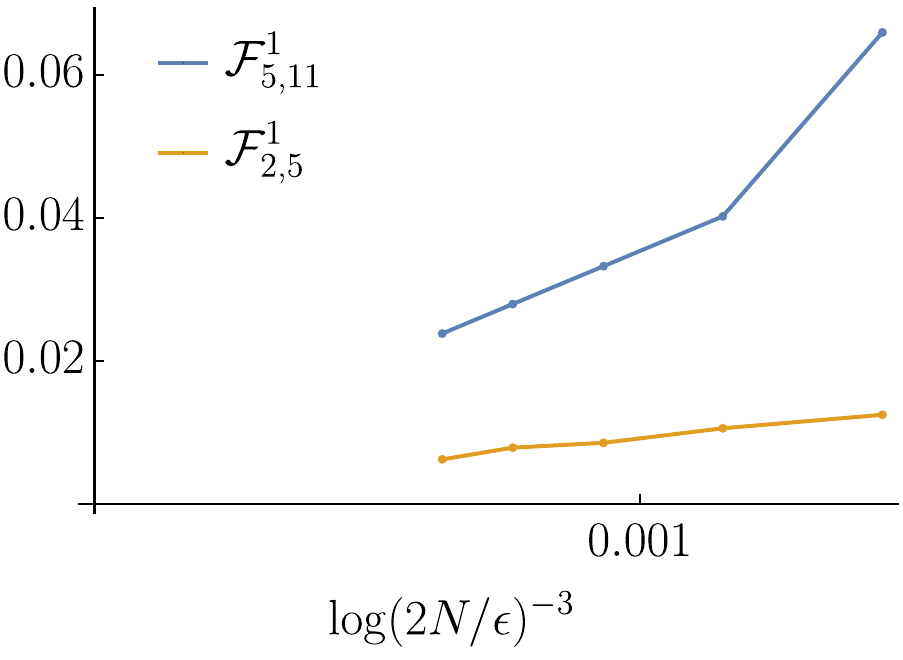}%
	\includegraphics[width=0.5\linewidth]{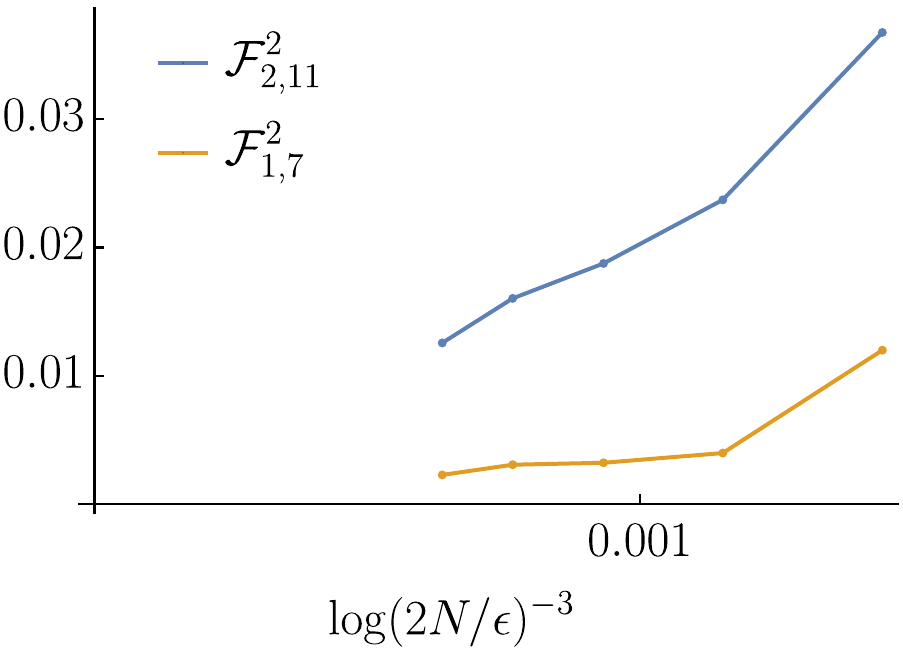}
	\caption{Estimate of $\ep$ by linear extrapolation and finite size corrections $\mathcal{F}^{n}_{\alpha\alpha'}$ for the XX model with interval size $2N$ = 512, 1024, 2048, 4096, 8192.
	}
	\label{fig:XX_Hn}
\end{figure}

There is a high degree of degeneracy because of the large symmetry group: the U(1) symmetry and $Z_2$ (spin flip around the X axis) symmetry. We can identify the Schmidt vectors with the scaling operators in the BCFT using the symmetry. For example, to identify the Schmidt vector corresponding to the stress tensor $L_{-2}\mathbb 1$, we find the Schimdt vector of scaling dimension close to 2, which is charge neutral under the U(1) symmetry and even under the $Z_2$ symmetry. After identifying the first few Schimdt vectors, we are ready to compute the matrix elements of $H_n$, which are also shown to converge to the exact values, see Fig. \ref{fig:XX_Hn}. 

\end{document}